\documentclass[12pt,a4paper]{iopart}
\pdfoutput=1
\usepackage[utf8]{inputenc}

\usepackage{graphicx}

\makeatletter
\providecommand*\@nameundef[1]{\expandafter\let\csname #1\endcsname\@undefined}
\@nameundef{equation*}
\@nameundef{endequation*}
\makeatother
\usepackage{amsmath}
\usepackage{amssymb}
\usepackage{amsfonts}
\usepackage{mathtools}
\usepackage{lmodern}
\usepackage[T1]{fontenc}
\usepackage{bbm}
\DeclareMathAlphabet{\mathbfi}{OML}{cmm}{b}{it}

\usepackage[unicode=true,bookmarksopen=true,colorlinks=true,urlcolor=blue,anchorcolor=blue,citecolor=blue,filecolor=blue,linkcolor=blue,menucolor=blue,
linktocpage=true,pdfa=true]{hyperref}

\let\originalleft\left
\let\originalright\right
\renewcommand{\left}{\mathopen{}\mathclose\bgroup\originalleft}
\renewcommand{\right}{\aftergroup\egroup\originalright}

\makeatletter
\newenvironment{equations}[1][]{\subequations\ifx\relax#1\relax\else\label{#1}\fi\align\ignorespaces}{\endalign\ignorespacesafterend\endsubequations}
\def\@spliteq#1{\begin{equation}\begin{split}#1\end{split}\end{equation}}
\def\splitequation{\collect@body\@spliteq}

\makeatother

\renewcommand{\vec}[1]{{\ifnum9<1#1\mathbf{#1}\else\ifcat\noexpand#1\relax\boldsymbol{#1}\else\mathbfi{#1}\fi\fi}}
\newcommand{\mathe}{\mathrm{e}}
\newcommand{\mathi}{\mathrm{i}}
\let\oldre\Re
\let\oldim\Im
\renewcommand{\Re}{\oldre\mathfrak{e}\,}
\renewcommand{\Im}{\oldim\mathfrak{m}\,}

\newcommand{\del}{\partial}
\newcommand{\laplace}{\mathop{}\!\bigtriangleup}
\newcommand{\abs}[1]{{\left\lvert{#1}\right\rvert}}

\newcommand{\eqend}[1]{\,#1}
\newcommand{\bigo}[1]{\mathcal{O}\left({#1}\right)}

\newcommand{\expect}[1]{\left\langle{#1}\right\rangle}

\newcommand{\hankel}[2]{\mathop{}\!\mathrm{H}^{(#1)}_{#2}}

\newcommand{\GFH}{G_\mathrm{H}^\mathrm{F}(x,x')}
\newcommand{\GFQ}{G_\mathrm{Q}^\mathrm{F}(x,x')}
\newcommand{\DFH}{D_\mathrm{H}^\mathrm{F}(x,x')}
\newcommand{\DFQ}{D_\mathrm{Q}^\mathrm{F}(x,x')}

\usepackage[numbers,sort&compress]{natbib}
\allowdisplaybreaks
\begin{document}

\title{Graviton backreaction on the local cosmological expansion in slow-roll inflation}

\author{William C. C. Lima}
\address{Department of Mathematics, University of York, Heslington, York, YO10 5DD, United Kingdom}
\ead{\href{mailto:william.correadelima@york.ac.uk}{william.correadelima@york.ac.uk}}

\begin{abstract}
We compute the graviton one-loop correction to the expectation value of the local expansion rate in slow-roll inflation, with both slow-roll parameters finite. The calculation is based on a recent method to explicitly construct gauge-invariant observables in perturbative quantum gravity at all orders in perturbation theory, and it is particularly suited in cases of highly-symmetrical space-time backgrounds. Our analysis adds to recent calculations of that correction in de Sitter space-time and in single-field inflation with constant deceleration. In the former case a vanishing one-loop correction was found, while in the latter the quantum backreaction produces a secular effect that accelerates the expansion. The quantum correction we describe here produces a finite secular effect that can either accelerated or decelerate the background expansion, depending on the value of the slow-roll parameters. 

\noindent\textit{Keywords}: perturbative quantum gravity, invariant observables, expansion rate, Hubble rate, inflation
\end{abstract}

\pacs{04.62.+v, 04.60.Bc, 11.15.-q, 04.60.-m}
% 04.62.+v	Quantum fields in curved spacetime
% 04.60.-m	Quantum gravity
% 04.60.Bc	Phenomenology of quantum gravity
% 11.10.Lm	Nonlinear or nonlocal theories and models
% 11.15.-q	Gauge field theories

\maketitle

%%%%%%%%%%%%%%%%%%%%%%%%%%%%%%%%%%%%%%%%%%%%%%%%%%%%%%%%%%%%%%%%%%%%%%%%%%%%%%%%%%%%%%%%%%%%%%%%%%%%%%%%%%%%%%%%%%%%%%%%%%%%%%%%%%%%%%%%%%%%%%%%%%%%%
\section{Introduction}                                                                                                                              %
\label{sec:introduction}                                                                                                                            %
%%%%%%%%%%%%%%%%%%%%%%%%%%%%%%%%%%%%%%%%%%%%%%%%%%%%%%%%%%%%%%%%%%%%%%%%%%%%%%%%%%%%%%%%%%%%%%%%%%%%%%%%%%%%%%%%%%%%%%%%%%%%%%%%%%%%%%%%%%%%%%%%%%%%%

According to our current understanding of the evolution of the early universe, the anisotropies in the cosmological microwave background (CMB)~\cite{planck_2018a, planck_2018b, planck_2018c} were created by quantum fluctuations of the metric during the inflationary era~\cite{guth1981, sato1981, linde1982, albrechtsteinhardt1982}. The picture drawn from treating gravity as an effective quantum field theory at energies below the Planck scale~\cite{burgess_llr_2004}, an approach known as perturbative quantum gravity, is the following. The fast space-time expansion during inflation copiously excites gravitons out of the vacuum~\cite{grishchuk_zetf_1974, ford_parker_prd_1977}. At tree level these quantum excitations created small fluctuations in the density profile of the cosmological fluid by gravitational attraction, which then show up in CMB maps today as regions slightly hotter or colder than the background~\cite{starobinskii_pzetf_1979, mukhanov_chibisov_pzetf_1981}.

Einstein's equation, however, is non-linear, and at higher loop orders the gravitons interact. Thus, it is only natural to ask what kind of effects this quantum fluctuations can have over the space-time evolution. In the case of a homogeneous and isotropic background, the question of what is the quantum gravitational backreaction on the cosmological expansion rate has received the attention of many authors over time---see e.g. Refs.~\cite{tsamis_woodard_npb_1996, mukhanov_abramo_brandenberger_prl_1997, unruh_arxiv_1998, abramo_woodard_prd_1999a, abramo_woodard_prd_1999b, abramo_woodard_prd_2002, geshnizjani_brandenberger_prd_2002, geshnizjani_brandenberger_jcap_2005, geshnizjani_afshordi_jcap_2005, losic_unruh_prd_2005, marozzi_vacca_cqg_2011, marozzi_vacca_brandenberger_jcap_2013, marozzi_vacca_prd_2014, miao_tsamis_woodard_prd_2017, froeb_cqg_2019}. In part, this was motivated by the possibility of backreaction effects producing a secular screening of the cosmological constant at one-loop order in de~Sitter space-time~\cite{ford_prd_1985, tsamis_woodard_plb_1993} and in Friedmann-Lema\^\i tre-Robertson-Walker (FLRW) space-times sourced by a scalar field (single-field inflation)~\cite{mukhanov_abramo_brandenberger_prl_1997}. The existence of these effects was questioned by Unruh~\cite{unruh_arxiv_1998}, who noted that local observers are not be able to measure them, which triggered a long debate on gauge issues related to the problem~\cite{abramo_woodard_prd_1999a, abramo_woodard_prd_1999b, abramo_woodard_prd_2002, geshnizjani_brandenberger_prd_2002, geshnizjani_brandenberger_jcap_2005, geshnizjani_afshordi_jcap_2005, losic_unruh_prd_2005, marozzi_vacca_cqg_2011, marozzi_vacca_brandenberger_jcap_2013, marozzi_vacca_prd_2014, miao_tsamis_woodard_prd_2017, froeb_cqg_2019, garriga_tanaka_prd_2007, tsamis_woodard_prd_2007, tsamis_woodard_prd_2013}.

At the heart of that debate lies the fact that the identification of gauge-invariant observables in general relativity that can be measured locally is a notoriously difficult task. Within the perturbative approach, this is mainly due to the fact that general relativity is a diffeomorphism invariant theory. This implies that no locally-defined scalar $S(x)$ can be gauge invariant unless it is a constant, since it will change at least with $\del_\mu S(x)$ under a diffeomorphism transformation. Thus, as noted by Torre~\cite{torre_prd_1993}, observables in general relativity are necessarily non-local---see also Refs.~\cite{giddings_marolf_hartle_prd_2006, khavkine_cqg_2015}. This is not the case of linearised gravity, where it is possible to find a complete set of local and gauge-invariant observables~\cite{froeb_hack_higuchi_jcap_2017, froeb_hack_khavkine_cqg_2018, khavkine_cqg_2019}, because any local quantity that vanishes on the background is gauge-invariant at linear order.

There are, however, proposals for how to obtain gauge-invariant observables that are non-local at both the perturbative and full non-linear regimes~\cite{tambornino_sigma_2012}. At the perturbative level, a method to construct such observables was recently put forward by Brunetti~{\it et al}~\cite{brunetti_et_al_jhep_2016} and further developed by Fr{\"o}b and Lima in Refs.~\cite{froeb_cqg_2018, froeb_lima_cqg_2018}. In this proposal the perturbed space-time points are labelled by field-dependent coordinates obtained as solutions of scalar differential equations. The field operators corresponding to the observables are then made gauge-invariant when expressed in terms of these coordinates, which depend non-locally on the space-time metric (and possibly other fields), and are of the relational type. This method is valid for general space-times, particularly in cases where the space-time background is highly symmetrical, such as cosmological space-times. An important aspect of this approach is that even though the observables are non-local, they are causal, i.e. they depend on the perturbations only in the space-time region within the past lightcone of the observation point. This is important, as it avoids potential unphysical ``action-at-a-distance'' processes~\cite{froeb_cqg_2018, froeb_lima_cqg_2018}. An alternative (but somewhat related) method to construct gauge-invariant observables is Dirac's ``dressing'' method, originally developed for quantum electrodynamics~\cite{dirac_cjp_1955} and more recently extended to perturbative quantum gravity~\cite{ware_saotome_akhoury_jhep_2013, donnelly_giddings_prd_2016a, donnelly_giddings_prd_2016b}.

Our aim in this paper is to make use of the proposal of Refs.~\cite{brunetti_et_al_jhep_2016,froeb_cqg_2018, froeb_lima_cqg_2018} to compute the quantum-gravitational one-loop correction to a gauge-invariant observable measuring the space-time local expansion rate in slow-roll inflation, with both slow-roll parameters finite. Similar calculations were recently performed in de~Sitter space-time~\cite{miao_tsamis_woodard_prd_2017} and in single-field inflation with constant deceleration parameter~\cite{froeb_cqg_2019}. The calculation of Miao~{\it et al}~\cite{miao_tsamis_woodard_prd_2017} has confirmed earlier expectations~\cite{abramo_woodard_prd_2002,geshnizjani_brandenberger_prd_2002} that there is no quantum-gravitational backreaction on the expansion rate in de~Sitter at one-loop order, although they leave the door open for effects at two-loop order. Shortly after Fr\"ob~\cite{froeb_cqg_2019} performed the calculation of the one-loop correction to the gauge-invariant local expansion rate as defined in Ref.~\cite{froeb_lima_cqg_2018} in single-field inflation with constant deceleration parameter, and found a secular backreaction effect that accelerates the expansion. 

With respect to inflationary scenarios, the constant deceleration case considered in Ref.~\cite{froeb_cqg_2019} corresponds to the power-law inflationary models~\cite{abbott_wise_npb_1984, lucchin_matarrese_prd_1985, sahni_cqg_1988}. These models have been discarded as viable inflationary models on the grounds that its prediction to the tensor-to-scalar ratio is in disagreement with the limits set by current experimental data~\cite{planck_2018c}. Slow-roll inflation, on the other hand, is a quite robust class of models~\cite{vieira_byrnes_lewis_jcap_2018} and remains as a viable candidate according to that data. Therefore, it is worthwhile to extend the calculation of the quantum-gravitational backreaction on the expansion rate to slow-roll inflation.

The paper is organised as follows. In Sec.~\ref{sec:gauge_invariant_observables} we revisit the proposal of Refs.~\cite{brunetti_et_al_jhep_2016, froeb_cqg_2018, froeb_lima_cqg_2018} in cosmological space-times and single-field inflation. In particular, we show how a simple generalisation in the definition of the configuration-dependent coordinates of Refs.~\cite{brunetti_et_al_jhep_2016,froeb_cqg_2018, froeb_lima_cqg_2018} can easily accommodate the method used in the de~Sitter calculation of Ref.~\cite{miao_tsamis_woodard_prd_2017}. In Sec.~\ref{sec:one_loop_correction} we introduce the observable describing the local cosmological expansion in single-field inflation, compute and renormalise its expectation value to one-loop order in slow-roll inflation. We present our conclusions in Sec.~\ref{sec:conclusions}. Some details of the calculations can be found in the appendices. We use the $-++\dots+$ convention for the metric signature in a $n$-dimensional space-time and set $c = \hbar = 1$ and $\kappa^2 \equiv 16\pi G_\mathrm{N}$.

%%%%%%%%%%%%%%%%%%%%%%%%%%%%%%%%%%%%%%%%%%%%%%%%%%%%%%%%%%%%%%%%%%%%%%%%%%%%%%%%%%%%%%%%%%%%%%%%%%%%%%%%%%%%%%%%%%%%%%%%%%%%%%%%%%%%%%%%%%%%%%%%%%%%%
\section{Gauge-invariant observables in cosmological space-times}                                                                                   %
\label{sec:gauge_invariant_observables}                                                                                                             %
%%%%%%%%%%%%%%%%%%%%%%%%%%%%%%%%%%%%%%%%%%%%%%%%%%%%%%%%%%%%%%%%%%%%%%%%%%%%%%%%%%%%%%%%%%%%%%%%%%%%%%%%%%%%%%%%%%%%%%%%%%%%%%%%%%%%%%%%%%%%%%%%%%%%%

Although coordinate systems are completely arbitrary, it is desirable to have them built out some notion of clock and rods. This is the case of the FLRW metric when written in terms of the co-moving coordinates:
\begin{equation}\label{eq:FLRW_metric_comoving}
 ds^2 = g_{\mu\nu}dx^\mu dx^\nu = -dt^2 + a^2(t)d\vec{x}^2\eqend{,}
\end{equation}
where $a$ is the scale factor. There, the clock is the cosmological fluid that sources that metric via Einstein's equation, and it is measured by the observers co-moving with it. The spatial coordinates are constructed out of the spatial distances defined by the Euclidean metric induced on the hypersurfaces where the fluid energy density is homogeneous and isotropic at an arbitrary time $t_0$, when $a(t_0) = 1$ by construction. Hence, all background observables are measured with respect to the cosmological fluid and the Euclidean metric induced on its contour hypersurfaces.

When considering perturbations on top of this space-time, we are forced to identify the points of the background with the ones of the full space-time. Part of those perturbations, the gauge part, correspond to a diffeomorphism from the background into itself, and can be interpreted as a ``small'' change of coordinates of the background. As a consequence, the Cartesian coordinates covering the FLRW background loose their nice physical meaning. Since observables are defined with respect to the clock and rods employed in their measurement, one cannot simply express the observables related to the full space-time in terms of the background coordinates. The usual way to go around this is to impose a gauge-fixing condition so to make that identification possible, and look for gauge-invariant combinations of the perturbations. However, this identification should not depend on a gauge-fixing condition to be implemented. Moreover, the task of finding gauge-invariant combinations of the perturbations becomes increasingly complex as one goes to orders higher than one in perturbation theory~\cite{acquaviva_et_al_npb_2003}, thus it is important to have an algorithm to obtain such combinations more systematically.

A way in those directions is via the relational approach. In this approach observables are obtained by considering the corresponding field at a point where other fields have prescribed values. These observables have been discussed in the general relativity literature starting in the 1950's~\cite{geheniau_debever_barb_1956a, geheniau_barb_1956, debever_barb_1956a, debever_barb_1956b, geheniau_debever_barb_1956b, komar_pr_1958, bergmann_komar_prl_1960, bergmann_rmp_1961}, and have been developed since, see e.g. Ref.~\cite{tambornino_sigma_2012} for a recent review. The relational approach relies on the construction of scalar fields as functionals of the fields $\psi$ in the system. These scalars are then employed as configuration-dependent coordinates $\tilde{X}^{(\alpha)}[\psi]$, with $\alpha = 0,1,\dots,n - 1$. To obtain such coordinates one can rely e.g.\ on geometrical scalars constructed out of the metric and its derivatives. Their viability as coordinates, however, depends on the background space-time to be generic enough, such that one is able to differentiate points by the values of these scalars. Another way is to simply introduce the scalars by hand, such as in the case of the Gaussian~\cite{kuchar_torre_prd_1991} and Brown-Kucha\v{r}~\cite{brown_kuchar_prd_1995} dust models. This, however, changes the physical content of the theory and can affect the observables, as shown by Giesel {\it et al}~\cite{giesel_et_al_cqg_2010a, giesel_et_al_cqg_2010b, giesel_et_al_arxiv_2020}.

\subsection{Configuration-dependent coordinates}
%%%%%%%%%%%%%%%%%%%%%%%%%%%%%%%%%%%%%%%%%%%%%%%%%%%%%%%%%%%%%%%%%%%%%%%%%%%%%%%%%%%%%%%%%%%%%%%%%%%%%%%%%%%%%%%%%%%%%%%%%%%%%%%%%%%%%%%%%%%%%%%%%%%%%

The problem of building relational observables in highly symmetrical geometries, such as FLRW space-times, has been overcome only recently in Ref.~\cite{brunetti_et_al_jhep_2016}. The solution presented there is to (perturbatively) construct the configuration-dependent scalars from scalar differential equations that are known to be satisfied on the background. For perturbations around Minkowski space-time in Cartesian coordinates, for instance, a simple choice is to take~\cite{froeb_cqg_2018}
\begin{equation}\label{eq:harmonic_coordinates}
 \tilde{\nabla}^2\tilde{X}^{(\alpha)}[\tilde{g}] = 0\eqend{,}
\end{equation}
where $\tilde{\nabla}^2 \equiv \tilde{\nabla}^\mu\tilde{\nabla}_\mu$ denotes the Laplace-Beltrami operator of the perturbed metric $\tilde{g}_{\mu\nu}$. Note that the coordinates $\tilde{X}^{(\alpha)}$ are scalars, therefore the notation with the index $\alpha$ within parenthesis. In fact, the coordinates defined by Eq.~(\ref{eq:harmonic_coordinates}) can be employed in perturbed space-times around arbitrary backgrounds, as long as the background is covered by coordinates satisfying the wave equation. 

That equation can be generalised in different ways. A possibility is to consider
\begin{equation}\label{eq:general_coordinates}
 \tilde{\nabla}^2\tilde{X}^{(\alpha)}[\tilde{g}] = F^{(\alpha)}(\tilde{X})\eqend{.}
\end{equation}
Since we want this equation to be fulfilled on the background, we choose
\begin{equation}
 F^{(\alpha)}(\tilde{X}) = (\nabla^2 x^\alpha)(\tilde{X})\eqend{,}
\end{equation}
where $x^\alpha$ denotes the background coordinates and $(\nabla^2 x^\alpha)(\tilde{X})$ means we replace $x^\alpha$ by $\tilde{X}^{(\alpha)}$ after we have computed the derivative, i.e. we keep the functional form of the result. As an example, let us consider the case of perturbations around a pure de~Sitter space-time and cover that background with the usual co-moving coordinates, i.e. we take $a(t) = \mathe^{Ht}$ in Eq.~(\ref{eq:FLRW_metric_comoving}), with $H$ as the Hubble constant. Then, $\nabla^2t = -(n - 1)H$ and $\nabla^2x^i = 0$, and thus Eq.~(\ref{eq:general_coordinates}) defines the configuration-dependent coordinates $\tilde{X}^{(\alpha)}$ in the perturbed space-time as
\begin{equation}
 \tilde{\nabla}^2\tilde{X}^{(0)}[\tilde{g}] = - (n - 1)H\eqend{,}\;\;\; \tilde{\nabla}^2\tilde{X}^{(i)}[\tilde{g}] = 0\eqend{.}
\end{equation}
Apart from an overall sign, $\tilde{X}^{(0)}$ above is precisely the non-local scalar field used by Tsamis and Woodard~\cite{tsamis_woodard_prd_2013} in their definition of an observable accounting for the local expansion on de~Sitter background, and later used by Miao~{\it et al}~\cite{miao_tsamis_woodard_prd_2017} in their loop calculation. Their observable, however, is invariant only with respect to pure time coordinate transformations, and thus useful in more restrict context where both the background and the state of perturbations are spatially homogeneous.

In what follows we will be interested in single-field inflationary models, where a spatially flat FLRW space-time is sourced by a scalar degree of freedom $\phi$, the inflaton. In this context it is more convenient to work with the FLRW background in conformally flat coordinates:
\begin{equation}
 ds^2 = a^2(\eta)\left(-d\eta^2 + d\vec{x}^2\right)\eqend{,}
\end{equation}
where $\eta$ is the conformal time. We also assume that the gradient of $\phi$ is everywhere time-like, with the derivative with respect to the conformal time as $\phi' < 0$, and that the metric and the scalar field satisfy Einstein's equation sourced by the scalar field with a scalar potential $V(\phi)$. This last assumption leads to the Friedmann equations
\begin{equations}[eq:friedmann_eqs]
 \kappa^2 V(\phi) & = 2(n - 2)(n - 1 - \epsilon)H^2\eqend{,}\\
 \kappa^2 (\phi')^2 & = 2(n - 2) H^2 a^2\epsilon\eqend{.}
\end{equations}
The Hubble parameter $H$ and the first two slow-roll parameters $\epsilon$ and $\delta$ are defined from the scale factor according to\footnote{The slow-roll parameters defined in Eq.~(\ref{eq:slow_roll_parameters}) are related to the widely used Hubble slow-roll parameters $\epsilon_H$ and $\eta_H$ as $\epsilon = \epsilon_H$ and $\delta = \epsilon - \eta_H$, see e.g. Ref.~\cite{liddle_parsons_barrow_prd_1994}.}
\begin{equation}\label{eq:slow_roll_parameters}
  H \equiv \frac{a'}{a^2}\eqend{,}\;\;\; \epsilon \equiv - \frac{H'}{H^2a}\eqend{,}\;\;\; \delta \equiv \frac{\epsilon'}{2Ha\epsilon}\eqend{.}
\end{equation}
The background scalar field equation is obtained by taking the time derivative of the second Friedmann equation, resulting in
\begin{equation}\label{eq:scalar_field_eq}
 \phi'' = (1 - \epsilon + \delta)Ha\phi'\eqend{.}
\end{equation}
This equation will be useful in what follows.

We can then add perturbations on top of that background and write the metric for the resulting space-time as
\begin{equation}\label{eq:perturbations}
 g_{\mu\nu} \to \tilde{g}_{\mu\nu} = a^2(\eta_{\mu\nu} + \kappa h_{\mu\nu})\qquad \mathrm{and}\qquad \phi \to \tilde{\phi} = \phi + \kappa \phi^{(1)}\eqend{.} 
\end{equation}
In single-field inflationary models the system provides a natural choice for a clock, namely, the inflaton~\cite{geshnizjani_brandenberger_prd_2002}. Thus, instead of relying on Eq.~(\ref{eq:general_coordinates}) to define a configuration-dependent time coordinate in the perturbed space-time, here we define $\tilde{X}^{(0)}$ by inverting the background relation $\phi = \phi(\eta)$ and evaluating it for $\tilde{\phi}$. That is, here we define the following local configuration-dependent time coordinate:
\begin{equation}\label{eq:time_single_field_inflation}
 \tilde{X}^{(0)}(x) \equiv \eta[\tilde{\phi}(x)]\eqend{.}
\end{equation}
The background spatial coordinates satisfy $\nabla^2 x^i = 0$, therefore from Eq.~(\ref{eq:general_coordinates}) we define the coordinates $\tilde{X}^{(i)}$ as
\begin{equation}\label{eq:spatial_harmonic_coordinates}
 \tilde{\nabla}^2 \tilde{X}^{(i)}[\tilde{g}] = 0\eqend{.}
\end{equation}
We remark that Eq.~(\ref{eq:spatial_harmonic_coordinates}) allows for different choices of Green's functions consistent with different choices of initial conditions, each of which corresponding to a different definition of the coordinates $\tilde{X}^{(\alpha)}$. Since we are interested in the causal evolution of the observables expectation value, the natural choice in that case is to impose the initial conditions $\tilde{X}^{(i)}(\eta_0,\vec{x}) = x^i$ and $\del_t\tilde{X}^{(i)}(\eta_0,\vec{x}) = 0$ and use the retarded Green's function. For further discussions on this point, see Refs.~\cite{froeb_cqg_2019,froeb_lima_cqg_2018}. 

Next, we expand the configuration-dependent coordinates $\tilde{X}^{(\alpha)}$ as
\begin{equation}\label{eq:expansion_X_tilde}
 \tilde{X}^{(\alpha)} = x^\alpha + \sum_{\ell = 1}^\infty \kappa^\ell X^{(\alpha)}_{(\ell)}(x)\eqend{.}
\end{equation}
The perturbative expansion of $\tilde{X}^{(0)}$ is simple to obtain, and up to second order in the perturbations it yields
\begin{equations}[eq:expansion_X_tilde_0]
 & X^{(0)}_{(0)}(x) = \eta\eqend{,}\\
 &\nonumber\\
 & X^{(0)}_{(1)}(x) = \frac{\del\eta}{\del\phi}[\phi(\eta)]\phi^{(1)}(x) = \frac{\phi^{(1)}(x)}{\phi'}\eqend{,}\\
 &\nonumber\\
 & X^{(0)}_{(2)}(x) = \frac{1}{2}\frac{\del^2\eta}{\del\phi^2}[\phi(\eta)]\left[\phi^{(1)}(x)\right]^2 = -\frac{\phi''}{2(\phi')^3}\left[\phi^{(1)}(x)\right]^2 = -\frac{(1 - \epsilon + \delta)Ha}{2(\phi')^2}\left[\phi^{(1)}(x)\right]^2\eqend{,}
\end{equations}
where we have used Eq.~(\ref{eq:scalar_field_eq}) in the last equation. At first order, we have for Eq.~(\ref{eq:spatial_harmonic_coordinates}) that~\cite{froeb_lima_cqg_2018} 
\begin{equation}
 \left[\del^2 - (n - 2)(Ha)(\eta)\del_\eta\right]X^{(i)}_{(1)}(x) = \del_\nu h^{i\nu}(x) - \frac{1}{2}\del^i h(x) + (n - 2)(Ha)(\eta) h^{0i}(x),
\end{equation}
 where $\del^2 \equiv \del^\alpha\del_\alpha$ and $h \equiv \eta^{\mu\nu}h_{\mu\nu}$, and with the initial conditions $X^{(i)}_{(1)}(\eta_0,\vec{x}) = 0$ and $\del_\eta X^{(i)}_{(1)}(\eta_0, \vec{x}) = 0$. Hence, the solution for this equation is
 \begin{equation}\label{eq:X_i_1}
  X^{(i)}_{(1)}(x) = \int d^nx' a^{n - 2}(\eta')G^\mathrm{ret}_\mathrm{H}(x,x')\left[\del_\nu h^{i\nu}(x') - \frac{1}{2}\del^i h(x') + (n - 2)(Ha)(\eta') h^{0i}(x')\right]\eqend{.}
 \end{equation}
In the expression above, $G^\mathrm{ret}_\mathrm{H}$ is the retarded Green's function defined in Ref.~\cite{froeb_lima_cqg_2018} and it satisfies
\begin{equation}
 \left[\del^2 - (n - 2)(Ha)(\eta)\del_\eta\right]G^\mathrm{ret}_\mathrm{H}(x,x') = \frac{1}{a^{n - 2}(\eta)}\delta^{(n)}(x - x')\eqend{.}
\end{equation}
It is clear from Eq.~(\ref{eq:X_i_1}) that the configuration-dependent spatial coordinates are non-local with respect to the metric perturbations. This non-locality is causal since the support of the retarded Green's function corresponds to the past light cone of the observation point $x$.

\subsection{Relational gauge-invariant observables}
%%%%%%%%%%%%%%%%%%%%%%%%%%%%%%%%%%%%%%%%%%%%%%%%%%%%%%%%%%%%%%%%%%%%%%%%%%%%%%%%%%%%%%%%%%%%%%%%%%%%%%%%%%%%%%%%%%%%%%%%%%%%%%%%%%%%%%%%%%%%%%%%%%%%%

Once the coordinates $\tilde{X}^{(\alpha)}$ have been defined, we can construct gauge-invariant observables in the following way. Consider a tensor $\tilde{T}^{\alpha_1\dots \alpha_k}_{\beta_1\dots\beta_m}$ on the perturbed space-time, which is constructed from the space-time metric and the inflaton. That quantity is made into a gauge-invariant observable by evaluating it at the point $x^\alpha$ corresponding to holding $\tilde{X}^{(\alpha)}$ fixed. This can be done simply by transforming its components to the coordinates $\tilde{X}^{\alpha}$:
\begin{equation}\label{eq:gauge_invariant_tensor}
 \mathcal{T}^{\mu_1\dots\mu_k}_{\nu_1\dots\nu_m}(\tilde{X}) \equiv \frac{\del \tilde{X}^{\mu_1}}{\del x^{\alpha_1}}\dots \frac{\del \tilde{X}^{\mu_k}}{\del x^{\alpha_k}} \frac{\del x^{\beta_1}}{\del \tilde{X}^{\nu_1}}\dots \frac{\del x^{\beta_m}}{\del \tilde{X}^{\nu_m}} \tilde{T}^{\alpha_1\dots \alpha_k}_{\beta_1\dots\beta_m}\left[x(\tilde{X})\right]{\Big{|}}_{\tilde{X}\;\textrm{fixed}}\eqend{,}
\end{equation}
where $x^\alpha(\tilde{X})$ denotes the inverse of $X^{(\alpha)}(x)$. It is interesting to notice that if we consider the metric perturbation $h_{\mu\nu}$ and take $X^{(\alpha)}$ as in Eqs.~(\ref{eq:time_single_field_inflation}) and~(\ref{eq:spatial_harmonic_coordinates}), then e.g.\ the time-time component of the corresponding gauge-invariant tensor, $\mathcal{H}_{00}$, is the Mukhanov-Sasaki variable at linear order~\cite{froeb_cqg_2019}.

As an useful example, let us consider Eq.~(\ref{eq:gauge_invariant_tensor}) when our observable is a scalar $S$. The perturbed scalar $\tilde{S}$ can be expanded as
\begin{equation}\label{eq:expansion_S}
 \tilde{S}(x) = \sum_{\ell = 0}^\infty\kappa^\ell S_{(\ell)}(x)\eqend{,}
\end{equation}
with $S_{(0)} = S$ as its background value. To obtain the perturbative expansion for $x^\alpha(\tilde{X})$, we need to invert the relation~(\ref{eq:expansion_X_tilde}). This can be easily done up to second order in $\kappa$:
\begin{splitequation}\label{eq:x_functional_X_tilde}
 x^\alpha 
 & = \tilde{X}^{(\alpha)} - \kappa X^{(\alpha)}_{(1)}(x) - \kappa^2 X^{(\alpha)}_{(2)}(x) + \dots\\
 & = \tilde{X}^{(\alpha)} - \kappa X^{(\alpha)}_{(1)}(\tilde{X} - \kappa X_{(1)}) - \kappa^2 X^{(\alpha)}_{(2)}(\tilde{X}) + \dots\\
 & = \tilde{X}^{(\alpha)} - \kappa X^{(\alpha)}_{(1)}(\tilde{X}) - \kappa^2\left[X^{(\alpha)}_{(2)}(\tilde{X}) - X^{(\sigma)}_{(1)}(\tilde{X})
 \del_\sigma X^{(\alpha)}_{(1)}(\tilde{X})\right] + \dots \eqend{.}
\end{splitequation}
By combining Eqs.~(\ref{eq:expansion_S}) and~(\ref{eq:x_functional_X_tilde}), we can then express the gauge-invariant observable corresponding to $S$ as
\begin{splitequation}\label{eq:scalar_observable}
 \mathcal{S}(\tilde{X}) 
 & \equiv \tilde{S}[x(\tilde{X})]\\
 & = S(\tilde{X}) + \kappa\left[S_{(1)}(\tilde{X}) - X^{(\sigma)}_{(1)}(\tilde{X})\del_\sigma S(\tilde{X})\right]\\ 
 & \phantom{=} + \kappa^2 \left[S_{(2)}(\tilde{X}) - X^{(\sigma)}_{(1)}(\tilde{X})\del_\sigma S_{(1)}(\tilde{X}) + \frac{1}{2}X^{(\rho)}_{(1)}(\tilde{X})X^{(\sigma)}_{(1)}(\tilde{X})\del_\rho\del_\sigma S(\tilde{X})\right. \\
 & \phantom{=} \left.\phantom{\frac{1}{2}} + X^{(\rho)}_{(1)}(\tilde{X})\del_\rho X^{(\sigma)}_{(1)}(\tilde{X})\del_\sigma S(\tilde{X}) - X^{(\sigma)}_{(2)}(\tilde{X})\del_\sigma S(\tilde{X})\right] + \dots\eqend{.}
\end{splitequation}
At this point we must refrain from using the relation $\tilde{X}^{(\alpha)} = \tilde{X}^{(\alpha)}(x)$ in the arguments of the functions appearing in the expression above, as that would send us back to the coordinate system $x^\alpha$. The coordinates now covering our space-time are $\tilde{X}^{(\alpha)}$, and since coordinates are mere labels, we can denote them by $x^\alpha$.

We remark that Eq.~(\ref{eq:gauge_invariant_tensor}) is invariant under diffeormorphisms that preserve the background fields by construction~\cite{brunetti_et_al_jhep_2016}. Nevertheless, it is possible to explicitly check this for Eq.~(\ref{eq:scalar_observable}) by using the gauge transformation of the inflaton and the metric perturbation and the definitions of the coordinates $\tilde{X}^{(\alpha)}$, and we refer the reader to Ref.~\cite{froeb_cqg_2019} for further details.

\subsection{The local expansion rate in single-field inflation}
%%%%%%%%%%%%%%%%%%%%%%%%%%%%%%%%%%%%%%%%%%%%%%%%%%%%%%%%%%%%%%%%%%%%%%%%%%%%%%%%%%%%%%%%%%%%%%%%%%%%%%%%%%%%%%%%%%%%%%%%%%%%%%%%%%%%%%%%%%%%%%%%%%%%%

We now turn our attention to a specific observable, the local expansion rate $H$, which measures the expansion of the space-time with respect to some notion of time. As mentioned above, in single-field inflation models it is natural to employ the inflaton field as our clock and in that case we can give a concrete definition of $H$ in terms of the divergence of the vector field normal to the contour hypersurfaces of $\phi$~\cite{geshnizjani_brandenberger_prd_2002}, i.e.
\begin{equation}\label{eq:local_Hubble_rate_definition}
 H \equiv \frac{\nabla^\mu u_\mu}{n - 1}\eqend{,}\;\;\; u_\mu \equiv \frac{\nabla_\mu\phi}{\sqrt{-\nabla^\sigma\phi\nabla_\sigma\phi}}\eqend{.}
\end{equation}

In the perturbed space-time we write $\tilde{H}$ in terms of the full metric and full inflaton and expand its expression up to second order in the perturbations. The result is
\begin{equation}
 \tilde{H}(x) = \frac{\tilde{\nabla}^\mu\tilde{u}_\mu}{n - 1} = H(\eta) + \kappa H^{(1)}(x) + \kappa^2 H^{(2)}(x) + \bigo{\kappa^3}\eqend{,}
\end{equation}
where the first- and second-order terms are given by
\begin{equations}[eq:terms_expansion_H]
 H^{(1)} =&\ \frac{1}{2(n - 1)a}(\del_\eta h^k_{\phantom{k}k} - 2\del_i h^i_{\phantom{i}0}) + \frac{H}{2}h_{00} - \frac{\laplace\phi^{(1)}}{(n - 1)a\phi'}\eqend{,}\\
 H^{(2)} =&\ \frac{(n - 3 + 2\epsilon - 2\delta)H}{2(n - 1)\phi'^2}\del^i\phi^{(1)}\del_i\phi^{(1)} + \frac{1}{2(n - 1)\phi'a}\left[(2\del_jh^{ij} + \del^ih_{00} - \del^ih^k_{\phantom{k}k})\del_i\phi^{(1)}\phantom{\frac{1}{1}}\right.\nonumber\\
          &\left. + 2h^{ij}\del_i\del_j\phi^{(1)} + \frac{2}{\phi'}\del_\eta\left(\del^i\phi^{(1)}\del_i\phi^{(1)}\right) + \frac{(\phi'h_{00} + 2\del_\eta\phi^{(1)})\laplace\phi^{(1)}}{\phi'} \right]\nonumber\\
          & + \frac{H}{2}\left(\frac{3}{4}h_{00}^2 - h_0^{\phantom{0}i}h_{i0}\right) - \frac{1}{4(n - 1)a}[2h_{ij}(\del_\eta h^{ij} - 2\del^ih^j_{\phantom{j}0})
            + 2h_{i0}(\del^ih^k_{\phantom{k}k} - \del_jh^{ij})\nonumber\\ 
          & - h_{00}(\del_\eta h^k_{\phantom{k}k} - 2\del^ih_{i0})]\eqend{.}
\end{equations}
We note that the indices above are raised and lowered with the Minkowski metric $\eta_{\mu\nu}$ and that to arrive at expression for $H^{(2)}$ in Eq.~(\ref{eq:terms_expansion_H}) we have used Eq.~(\ref{eq:scalar_field_eq}).

We then employ the procedure described above to obtain a gauge-invariant expression for the local expansion rate in the perturbed space-time. Hence, we define~\cite{froeb_lima_cqg_2018}
\begin{equation}\label{eq:gauge_invariant_H}
 \mathcal{H}(\tilde{X}) \equiv \tilde{H}[x(\tilde{X})] 
\end{equation}
and expand the resulting expression up to second order in the perturbations as
\begin{equation}
 \mathcal{H} = \mathcal{H}^{(0)} + \kappa \mathcal{H}^{(1)} + \kappa^2 \mathcal{H}^{(2)} + \bigo{\kappa^3}\eqend{.}
\end{equation}
From Eq.~(\ref{eq:scalar_observable}) we have that
\begin{equations}[eq:terms_expansion_gauge_invariant_H]
 \mathcal{H}^{(0)} =&\ H\eqend{,}\\
 \mathcal{H}^{(1)} =&\ H^{(1)} + H^2a\epsilon X^{(0)}_{(1)}\eqend{,}\\
 \mathcal{H}^{(2)} =&\ H^{(2)} - X^{(\mu)}_{(1)}\del_\mu H^{(1)} - \frac{1}{2}H^3a^2\epsilon(1 - 2\epsilon + 2\delta) X^{(0)2}_{(1)}\nonumber\\
                    &\ - H^2a\epsilon X^{(\mu)}_{(1)}\del_\mu X^{(0)}_{(1)} + H^2a\epsilon X^{(0)}_{(2)}\eqend{.}
\end{equations}
Hence, only the expression of the $\tilde{X}^{(0)}$ coordinate is needed to second order in $\kappa$. We remind the reader that, from the definition of the time coordinate $\tilde{X}^{(0)}$, $\mathcal{H}$ is measured with respect to a family of observers co-moving with the full inflation field $\tilde{\phi}$.

Note that up to this point we have an identification between the points of the background and the full space-times without the need of fixing the gauge and a truly gauge-invariant observable, $\mathcal{H}$, that is (necessarily) non-local. We can  now use the gauge freedom to simplify the forthcoming calculation. An obvious choice is to fix the gauge in order to have $\tilde{X}_{(1)}^{(\mu)} = 0$. To make the first-order perturbations of the coordinates $\tilde{X}^{(\mu)}$ to vanish amounts to imposing the following conditions exactly on the inflaton field and metric perturbations~\cite{froeb_lima_cqg_2018}:
\begin{equation}\label{eq:gauge_conditions}
 \phi^{(1)} = 0\eqend{,}\;\;\; \del_\mu h^{\mu i} - \frac{1}{2}\del^ih + (n - 2)Hah^{i0} = 0\eqend{.}
\end{equation}
The condition on the scalar field perturbation also implies that $\tilde{X}_{(2)}^{(0)} = 0$, as can be seen from Eq.~(\ref{eq:expansion_X_tilde_0}), and this gauge choice simplifies Eq.~(\ref{eq:terms_expansion_gauge_invariant_H}) to
\begin{equation}\label{eq:terms_expansion_gauge_invariant_H_simplified}
 \mathcal{H}^{(0)} = H\eqend{,}\;\;\; \mathcal{H}^{(1)} = H^{(1)}\eqend{,}\;\;\; \mathcal{H}^{(2)} = H^{(2)}\eqend{.}
\end{equation}
We remark that although the gauge-fixed expression for $\mathcal{H}$ given in Eq.~(\ref{eq:terms_expansion_gauge_invariant_H_simplified}) is local up to second order, this would not be the case had we chosen a different gauge-fixing condition.

%%%%%%%%%%%%%%%%%%%%%%%%%%%%%%%%%%%%%%%%%%%%%%%%%%%%%%%%%%%%%%%%%%%%%%%%%%%%%%%%%%%%%%%%%%%%%%%%%%%%%%%%%%%%%%%%%%%%%%%%%%%%%%%%%%%%%%%%%%%%%%%%%%%%%
\section{Graviton one-loop correction to the local expansion rate}                                                                                  %
\label{sec:one_loop_correction}                                                                                                                     %
%%%%%%%%%%%%%%%%%%%%%%%%%%%%%%%%%%%%%%%%%%%%%%%%%%%%%%%%%%%%%%%%%%%%%%%%%%%%%%%%%%%%%%%%%%%%%%%%%%%%%%%%%%%%%%%%%%%%%%%%%%%%%%%%%%%%%%%%%%%%%%%%%%%%%

In this section we calculate the expectation value of $\mathcal{H}$ up to one-loop order via the in-in (or closed-time path) formalism of Schwinger and Keldysh~\cite{schwinger_jmp_1961,keldysh_zetf_1964,chou_et_al_pr_1985}. This formalism computes true expectation values of the operators, rather than their matrix element between some in and out states, and is tailored for cases where the system evolves from a given initial state. This is enforced by integrating the vertices along a contour $C$ in the complex-$\eta$ plane, with a part $C_1$ that runs forward in (real) time from the initial time $\eta_0$ up to an arbitrary final time $\eta_\mathrm{f}$, larger than any external point time coordinate (assumed to be on $C_1$), and a part $C_2$ that runs backwards in time back to $\eta_0$. Because of this one has to use the contour-ordered propagator $G^\mathrm{c}$. At one-loop order, its integration along $C$ gives the Feynman propagator minus the negative-frequency Wightman two-point function. For early use of the in-in formalism in curved space-times, see e.g. Refs.~\cite{kay_cmp_1980, jordan_prd_1986, calzetta_hu_prd_1987}.

For finite initial time we must consider a dressed state for the interacting theory, in principle. A natural choice for the state of the interacting field, however, is to assume that in the asymptotic past its fluctuations are in  the free vacuum state and that the interaction is switched on adiabatically. In the case we are interested in, this choice for the initial state can be implemented just as in the Minkowski~\cite{itzykson_and_zuber_qft_book} or in the de~Sitter~\cite{adshead_easther_lim_prd_2009, frob_roura_verdaguer_jcap_2012} cases, by a time coordinate integration contour with an ever decreasing imaginary part and then taking $\eta_0 \to -\infty$. This is the well-known $\mathi \epsilon$ prescription. In the following-up calculation, however, the integrals converge even without the use of the $\mathi \epsilon$ prescription. 

The expansion rate expectation value up to one-loop order then reads
\begin{equation}\label{eq:H_expectation_value}
 \expect{\mathcal{H}(x)} = H + \mathi \kappa^2\expect{H^{(1)}(x)S^{(1)}_\textrm{int}}_0 + \mathi \kappa^2\expect{H^{(1)}(x)S^{(1)}_\textrm{G,CT}}_0 + \kappa^2\expect{H^{(2)}(x)}_0, 
\end{equation}
with the $\eta$ integration contour as described above. In this expression, $\langle\,\cdot\,\rangle_0$ denotes the expectation value with respect to the free-graviton state in the gauge given in Eq.~(\ref{eq:gauge_conditions}). The corresponding Feynman propagator
\begin{equation}
\mathi G^\mathrm{F}_{\mu\nu\rho\sigma}(x,x') \equiv \theta(\eta - \eta')G^+_{\mu\nu\rho\sigma}(x,x') + \theta(\eta' - \eta)G^+_{\mu\nu\rho\sigma}(x',x)\eqend{,}
\end{equation}
with the Heaviside step function $\theta(x)$ and the positive-frequency Wightman two-point function
\begin{equation}
\mathi G^+_{\mu\nu\rho\sigma}(x,x') \equiv \left\langle h_{\mu\nu}(x)h_{\rho\sigma}(x')\right\rangle_0\eqend{,}
\end{equation}
was constructed in Ref.~\cite{froeb_lima_cqg_2018}, and its expression is reproduced in \ref{apdx:free_propagators} for the reader's convenience. The term $S^{(1)}_\textrm{int} \equiv S^{(1)}_\textrm{GH,eff} + S^{(1)}_\textrm{G}$ in Eq.~(\ref{eq:H_expectation_value}) denotes the total interaction action, with $S^{(1)}_\textrm{GH,eff}$ the interaction of the graviton with the ghost fields and 
\begin{equation}
 S^{(1)}_\textrm{G} = S^{(1)}_\textrm{G,U} + S^{(1)}_\textrm{G,V}\eqend{,}
\end{equation}
with
\begin{equation}\label{eq:S_G_U}
 S^{(1)}_{\textrm{G},U} \equiv \frac{1}{8}U^{\alpha\beta\gamma\delta\mu\nu\rho\sigma}\int d^nxa^{n - 2}h_{\gamma\delta}\del_\alpha h_{\mu\nu}\del_\beta h_{\rho\sigma}
\end{equation}
and
\begin{equation}\label{eq:S_G_V}
 S^{(1)}_{\textrm{G},V} \equiv \frac{n - 2}{4}V^{\alpha\beta\mu\nu\rho\sigma}\int d^nxHa^{n - 1}h_{\alpha\beta}h_{0\sigma}\del_\rho h_{\mu\nu}\eqend{,}
\end{equation}
the three-gravitons interaction terms. $S^{(1)}_\textrm{G,CT}$ are the counter-terms necessary to absorb the divergences coming from the insertion of the basic fields. The explicit expressions for $S^{(1)}_\textrm{GH,eff}$ and $S^{(1)}_\textrm{G,CT}$ can be found in Ref.~\cite{froeb_cqg_2019}, and the form of the tensors $U^{\alpha\beta\gamma\delta\mu\nu\rho\sigma}$ and $V^{\alpha\beta\mu\nu\rho\sigma}$ are given in \ref{apdx:coincidence_limit_three_gravitons}.

\subsection{Slow-roll approximation}
%%%%%%%%%%%%%%%%%%%%%%%%%%%%%%%%%%%%%%%%%%%%%%%%%%%%%%%%%%%%%%%%%%%%%%%%%%%%%%%%%%%%%%%%%%%%%%%%%%%%%%%%%%%%%%%%%%%%%%%%%%%%%%%%%%%%%%%%%%%%%%%%%%%%%

In the slow-roll approximation we assume that $\epsilon \ll 1$ and $\abs{\delta} \ll 1$, and only keep terms linear in the small parameters $\epsilon$ and $\delta$---see, e.g., Refs.~\cite{liddle_parsons_barrow_prd_1994,lidsey_et_al_rmp_1997, oikonomou_epl_2020}. The definition of the first slow-roll parameter, Eqs.~(\ref{eq:slow_roll_parameters}), implies that $\epsilon' = \bigo{\epsilon \delta}$. Hence, we can neglect $\epsilon'$, unless it appears multiplied by an inverse power of a small parameter. We assume that the same is true for $\delta'$. Within this approximation, the integration of Eqs.~(\ref{eq:slow_roll_parameters}) gives
\begin{equation}\label{eq:slow_roll_ha}
\epsilon = \epsilon_0 a^{2\delta} \eqend{,} \qquad H = H_0 a^{-\epsilon} \eqend{,} \qquad a = \left[ - (1-\epsilon) H_0 \eta \right]^{- \frac{1}{1-\epsilon}} \eqend{,}
\end{equation}
where $H_0$ and $\epsilon_0$ are the values of the Hubble and the first slow-roll parameters at the initial time $\eta_0$, respectively. In particular, we have
\begin{equation}\label{eq:slow_roll_ha_in_eta}
H a = - \frac{1}{(1-\epsilon) \eta}
\end{equation}
as in the constant-$\epsilon$ case. 

The slow-roll approximation is only valid for some limited range of conformal times $\eta$. Indeed, by expanding $\epsilon$ in powers of $\delta$ in Eq.~(\ref{eq:slow_roll_ha}), we obtain
\begin{equation}
\epsilon = \epsilon_0 \left[ 1 + 2 \delta \ln a + 2 \delta^2 \ln^2 a + \bigo{\delta^3} \right] \eqend{.}
\end{equation}
Clearly we must have $\abs{ \delta \ln a } \ll 1$ in order to neglect the third and all higher-order terms. A similar expansion of the expression for $H$ leads to the condition $\abs{ \epsilon \ln a } \ll 1$. That is, the approximation is valid for as long as the logarithm of the scale factor changes much less than $N = 1/\max\left( \abs{\delta}, \epsilon \right)$. As a consequence, the observation time $\eta$ and the initial time $\eta_0$ must not be more than $N$ e-folds apart for a given expression to be valid.\footnote{If one is not interested in the coordinate-space expressions, but only in the results in Fourier space, the approximation can be improved by taking $\epsilon$ and $\delta$ constant but different for each mode, namely at horizon crossing where $H a = \abs{\vec{p}}$; see, e.g., Ref.~\cite{lidsey_et_al_rmp_1997}. The condition $\abs{ \{ \delta, \epsilon \} \ln a} \ll 1$ is then unnecessary.} We shall return to this point below when employing the in-in formalism.

While performing this calculation, we realised that there are a few missing factors in the expressions for the Fourier amplitude of the scalar propagators in Ref.~\cite{froeb_lima_cqg_2018}, which were used in the constant-$\epsilon$ calculation of Ref.~\cite{froeb_cqg_2019}. Moreover, we have also found a computational mistake in that reference when considering the contribution coming from the part of the three-gravitons interaction involving the $V$-tensor. We have checked these issues with the author of Ref.~\cite{froeb_cqg_2019} and corrected the formulas affected by them. In \ref{apdx:free_propagators} we point out the correct formulas for the scalar propagators of Ref.~\cite{froeb_lima_cqg_2018} that are relevant to the slow-roll calculation. The corrected expressions for the one-loop correction to $\mathcal{H}$ in the constant-$\epsilon$ case are shown in \ref{apdx:cte_epsilon} for the reader's convenience. We remark that these corrections are purely computational, and do no affect in any way the main conclusions of Refs.~\cite{froeb_cqg_2019, froeb_lima_cqg_2018}.

\subsection{Loop computation}
%%%%%%%%%%%%%%%%%%%%%%%%%%%%%%%%%%%%%%%%%%%%%%%%%%%%%%%%%%%%%%%%%%%%%%%%%%%%%%%%%%%%%%%%%%%%%%%%%%%%%%%%%%%%%%%%%%%%%%%%%%%%%%%%%%%%%%%%%%%%%%%%%%%%%

\subsubsection{The \texorpdfstring{$H^{(2)}$}{pure second-order} term}
%%%%%%%%%%%%%%%%%%%%%%%%%%%%%%%%%%%%%%%%%%%%%%%%%%%%%%%%%%%%%%%%%%%%%%%%%%
The contribution from the term $H^{(2)}(x)$, which was given in Eq.~(\ref{eq:terms_expansion_H}), only involves the coincidence limit of the graviton propagator. Hence, we can regularise its expression first via the point-splitting method and then use the dimensional regularisation approach to perform the momentum integrals. The result is
\begin{splitequation}\label{eq:H_(2)_graviton_propagator}
 \expect{H^{(2)}(x)}_0 
 & = - \mathi\lim_{x' \to x} \left\{\frac{1}{4(n - 1)a(\eta)}\left[(\del_\eta + \del_{\eta'})G^{\textrm{F} ij}_{\phantom{\textrm{F} ij}ij}(x,x') - 4\del^iG^{\textrm{F} j}_{\phantom{\textrm{F}j}0ij}(x,x')\right.\right. \\
 &\phantom{=}\qquad\qquad + 2\del^iG^{\textrm{F}j}_{\phantom{\textrm{F} j}j0i}(x,x') - 4\del_jG^{\textrm{F}ij}_{\phantom{\textrm{F}ij}0i}(x,x') 
 - \del_\eta G^{\textrm{F}i}_{\phantom{\textrm{F}i}i000}(x,x') \\
 &\phantom{=}\qquad\qquad \left.\left. + 2\del_iG^{\textrm{F}i}_{\phantom{\textrm{F}i}000}(x,x')\right] - \frac{H(\eta)}{2}\left[\frac{3}{4}G_{0000}^\textrm{F}(x,x') - G^{\textrm{F} i}_{\phantom{\textrm{F} i}0i0}(x,x')\right]\right\}\eqend{.}
\end{splitequation}
We use the expressions for the graviton propagator, Eqs.~(\ref{eq:graviton_propagator}), and with the help of Eqs.~(\ref{eq:G_2_D_2_propagators}) and~(\ref{eq:G_H_D_H_relation}) we obtain
\begin{splitequation}\label{eq:H_(2)_scalar_propagators_slow_roll}
 \expect{H^{(2)}(x)}_0 
 & = -\mathi \lim_{x' \to x}\left\{\frac{(2n - 1)(n - 2) - 1 + (n - 1)(n - 2)\epsilon}{4(n - 2)a(\eta)}(\del_\eta + \del_{\eta'})\GFH\right. \\
 &\phantom{=} - \frac{1 + \epsilon}{2(Ha^2)(\eta)}(\laplace + \del_\eta\del_{\eta'})\GFH\\ 
 &\phantom{=} - \frac{H(\eta)}{4}\frac{(n - 3)(n^2 - 3n + 3) - (n - 1)(n - 2)^2\epsilon}{n - 2}\DFH\\
 &\phantom{=} + \frac{1}{8(n - 1)(H^2a^3)(\eta)}\left[2(n - 1)(H^2a^2)(\eta)(\del_\eta + \del_{\eta'})\GFQ\right.\\ 
 &\phantom{=} + (n - 1)(Ha)(\eta)\laplace \GFQ - (n - 1 + 2\epsilon)(Ha)(\eta)\del_\eta\del_{\eta'}\GFQ \\
 &\phantom{=} \left. \left. + (\del_\eta + \del_{\eta'})\laplace \GFQ\right] - \frac{\epsilon^2H(\eta)}{8}\DFQ\right\}\eqend{,}
\end{splitequation}
where $\laplace$ is the familiar Laplace operator in Euclidean space.

To compute the coincidence limits of the derivatives of the scalar propagators in the expression above, we will rely on their Fourier transforms given in Eqs.~(\ref{eq:GFH_Fourier_transform}) - (\ref{eq:Fourier_transform_D_Q}). 
Assuming that $\lim_{\eta' \to \eta}\theta(\eta - \eta') = \frac{1}{2}$, we calculate e.g.
\begin{splitequation}
 \mathi \lim_{x' \to x}(\del_\eta + \del_{\eta'})G^\textrm{F}_\textrm{H}(x,x') = 
 & -\frac{\pi}{4}[(1 - \epsilon)H]^{n - 2}\eta^{n - 1}\int\frac{d^{n - 1}p}{(2\pi)^{n - 1}}p\left[\hankel{1}{\mu - 1}(-p\eta)\hankel{2}{\mu}(-p\eta)\right.\\ 
 &\left. + \hankel{1}{\mu}(-p\eta)\hankel{2}{\mu - 1}(-p\eta)\right]\eqend{,}
\end{splitequation}
where $\hankel{1}{\alpha}(x)$ and $\hankel{2}{\alpha}(x)$ are the Hankel functions of first and second kinds~\cite{dlmf}, respectively, and order $\alpha$, $p \equiv \abs{\vec{p}}$, and the parameter $\mu$ depends on the slow-roll parameters as
\begin{equation}\label{eq:mu_definition}
 \mu = \frac{n - 1}{2} + \frac{n - 2}{2}\epsilon\eqend{.}
\end{equation}
Integrals of Hankel functions in this form will often appear in this calculation. Following Ref.~\cite{froeb_cqg_2019}, it is convenient to define the dimensionless integral
\begin{equation}\label{eq:J_integral}
 J_{k,\alpha,\beta} \equiv \frac{\pi}{8} \int\frac{d^{n - 1}q}{(2\pi)^{n - 1}}q^k\left[\hankel{1}{\alpha}(q)\hankel{2}{\beta}(q) + \hankel{1}{\beta}(q)\hankel{2}{\alpha}(q)\right]\eqend{.}
\end{equation}
The integral $J_{k,\alpha,\beta}$ is calculated and analysed in detail in \ref{apdx:analysis_integral_J}. We then use Eqs.~(\ref{eq:coincidence_limit_G_HQ_D_HQ}) in Eq.~(\ref{eq:H_(2)_scalar_propagators_slow_roll}) to obtain
\begin{splitequation}
 & \expect{H^{(2)}(x)}_0\\
 & = [(1 - \epsilon)H]^{n - 1}\left[\frac{(2n - 1)(n - 2) - 1 + (n - 1)(n - 2)\epsilon}{2(n - 2)}J_{1,\mu,\mu - 1} + \frac{1}{2}(J_{2,\mu - 1,\mu - 1} - J_{2,\mu,\mu})\right.\\ 
 &\phantom{=} - \frac{(n - 3)(n^2 - 3n + 3) + (n - 1)(n - 2)^2\epsilon}{4(n - 2)(1 - \epsilon)}J_{0,\mu - 1,\mu - 1} + \frac{4J_{1,\nu,\nu - 1} + (1 - \epsilon)J_{2,\nu,\nu}}{4(n - 2)\epsilon}\\
 &\phantom{=} \left. + \frac{n - 1 - (n - 3)\epsilon}{4(n - 1)(n - 2)\epsilon}J_{2,\nu - 1,\nu - 1} - \frac{1 - 2\epsilon}{2(n - 1)(n - 2)\epsilon}J_{3,\nu,\nu - 1} - \frac{\epsilon}{4(n - 2)}J_{+,\nu - 1,\nu - 1}\right]\eqend{.}
 \end{splitequation}
Next, we write the expectation value above in the form
\begin{equation}
 \expect{H^{(2)}(x)}_0 = -H^{n - 1}D_2(n,\epsilon,\delta)
\end{equation}
and use Eqs.~(\ref{eq:J_integral_values}) to obtain
\begin{splitequation}\label{eq:D_2}
 D_2(n,\epsilon,\delta) 
 & = \frac{A_\mu^{(n)}}{4(n - 2)}\frac{(1 - \epsilon)^{n - 2}}{\epsilon}\left[n(13 - 6n - 2n^2 + n^3)\epsilon + \bigo{\epsilon^2}\right]\\
 & \phantom{=} + \frac{A_\nu^{(n)}}{16(n - 2)}\frac{(1 - \epsilon)^{n - 2}}{\epsilon} \left[4n(n^2 + n - 6) + 2(8 + 2n + 3n^2 - 3n^3)\epsilon\right.\\
 &\phantom{=} \left. - 4(4 - 5n - n^2)\delta + \bigo{\epsilon\delta}\right]\eqend{.}
\end{splitequation}

\subsubsection{The counter-terms}
%%%%%%%%%%%%%%%%%%%%%%%%%%%%%%%%%%%%%%%%%%%%%%%%%%%%%%%%%%%%%%%%%%%%%%%%%%
The contribution coming from the counter-terms is given by
\begin{splitequation}\label{eq:CT_integral}
 & \mathi \expect{H^{(1)}(x)S^{(1)}_\mathrm{G,CT}}_0\\
 &\quad = \mathi \frac{(n - 2)}{2}\int d^nx'(H^2a^n)(\eta')\left\{[(n - 1 - \epsilon)\delta_V + \epsilon\delta_Z] \expect{H^{(1)}(x)h_{00}(x')}_0 \right.\\
 &\qquad\qquad  \phantom{=} \left. - [(n - 1 - \epsilon)\delta_V - \epsilon\delta_Z]\expect{H^{(1)}(x)h^k_{\phantom{k}k}(x')}_0\right\}\eqend{,}
\end{splitequation}
where $\delta_V$ and $\delta_Z$ are the scalar potential and the field strength counter-terms, respectively. We can use the form of $H^{(1)}(x)$ given in Eq.~(\ref{eq:terms_expansion_H}) to express the expectation values appearing in the integrand of Eq.~(\ref{eq:CT_integral}) in terms of the graviton propagator as
\begin{equation}
 \expect{H^{(1)}(x)h_{\mu\nu}(x')}_0 = \frac{\mathi}{2(n - 1)a(\eta)}F_{\mu\nu}(x,x')\eqend{,} 
\end{equation}
with the definition
\begin{equation}\label{eq:F_mu_nu}
 F_{\mu\nu}(x,x') \equiv \del_\eta G^{\mathrm{c}\,k}_{\phantom{\mathrm{c}\,k}k\mu\nu}(x,x') - 2\del_kG^{\mathrm{c}\,k}_{\phantom{\mathrm{c}\,k}0\mu\nu}(x,x') + (n - 1)(Ha)(\eta)G^\mathrm{c}_{00\mu\nu}(x,x') \eqend{.}
\end{equation}
With the aid of Eqs.~(\ref{eq:graviton_propagator}), the components of $F_{\mu\nu}$ relevant to this section can be expressed in terms of the scalar propagators as
\begin{equation}\label{eq:F_00_slow_roll}
 F_{00}(x,x') = \frac{1}{(Ha)(\eta')}\laplace\left(\frac{1}{(Ha)(\eta)}\del_{\eta'}G^\mathrm{c}_\mathrm{Q}(x,x') 
 - \epsilon(\eta) D^\mathrm{c}_\mathrm{Q}(x,x')\right)
\end{equation}
and
\begin{equation}\label{eq:F_ij_slow_roll}
 F_{ij}(x,x') = -\delta_{ij}\left(\epsilon(\eta)\del_\eta - \frac{\laplace}{(Ha)(\eta)}\right)G^\mathrm{c}_\mathrm{Q}(x,x')\eqend{.}
\end{equation}

The contribution from the counter-terms can then be cast as
\begin{splitequation}\label{eq:CT_integral_F_mu_nu}
  \mathi \expect{H^{(1)}(x)S^{(1)}_\mathrm{G,CT}}_0 &= \mathi \frac{(n - 2)}{2}\int d^nx'(H^2a^n)(\eta')\left\{[(n - 1)\delta_V - (\delta_V -\delta_Z)\epsilon(\eta')] F^{\phantom{0}}_{00}(x,x') \right.\\
 &\qquad\qquad \left. - [(n - 1)\delta_V - (\delta_V + \delta_Z)\epsilon(\eta')]F^k_{\phantom{k}k}(x,x')\right\}\eqend{.}
\end{splitequation}
The Laplacian operator in the expression for $F_{00}$ acts on $x$ and, thus, can be pulled out of the integral. Moreover, the spatial homogeneity of our state and space-time background implies that the integral on $x'$ must be independent from the spatial coordinates. Therefore, the integration of $F_{00}$ vanishes. The same reasoning is valid for the term containing the Laplacian operator in the expression for $F_{ij}$, and its contribution also vanishes. In conclusion, we have reduced Eq.~(\ref{eq:CT_integral_F_mu_nu}) to
\begin{splitequation}\label{eq:CT_integral_I_J}
 \mathi \expect{H^{(1)}(x) S_\mathrm{G,CT}^{(1)}}_0
 & = -\frac{(n - 2)\epsilon}{4a(\eta)}\int d^nx (H^2a^n)(\eta')[(n - 1)\delta_V - (\delta_V - \delta_Z)\epsilon(\eta')]\del_\eta G^\mathrm{c}_\mathrm{Q}(x,x')\\
 &= -\frac{(n - 2)\epsilon}{4a(\eta)}\left[(n - 1)\delta_VI_{2,0}(\eta) - (\delta_V - \delta_Z)I_{2,1}(\eta)\right]\eqend{,}
\end{splitequation}
where we have defined 
\begin{equation}\label{eq:I_m_alpha_limit}
 I_{m,\alpha}(\eta) \equiv \lim_{\eta_0 \to -\infty} I_{m,\alpha}(\eta, \eta_0)\eqend{,}
\end{equation}
with the integral
\begin{equation}\label{eq:I_m_alpha_integral}
 I_{m,\alpha}(\eta, \eta_0) \equiv \int_{\eta_0}^\eta d\eta'\int d^{n - 1}x(\epsilon^\alpha H^ma^n)(\eta')\del_\eta[G_\textrm{Q}^+(x,x') - G_\textrm{Q}^+(x',x)]\eqend{.}
\end{equation}
Although we are employing the in-in formalism, where we take $\eta_0 \to -\infty$ so the interaction is switched off adiabatically in the asymptotic past, the slow-roll approximation is only valid for a finite number of e-folds. We will discuss this conflict after performing the integral~(\ref{eq:I_m_alpha_integral}).

The integral $I_{m,\alpha}$ can be computed as follows. We consider the expression of the Wightman two-point function $G^+_\mathrm{Q}$ in terms of its Fourier amplitude $\tilde{G}^+_\mathrm{Q}$ and then integrate Eq.~(\ref{eq:I_m_alpha_integral}) over the spatial coordinates. The result is
\begin{equation}\label{eq:I_m_alpha_Fourier}
 I_{m,\alpha}(\eta) = \int_{\eta_0}^\eta d\eta'(\epsilon^\alpha H^ma^n)(\eta')\del_\eta\lim_{\vec{p} \to 0}[\tilde{G}^+_\mathrm{Q}(\eta,\eta',\vec{p}) - \tilde{G}^+_\mathrm{Q}(\eta',\eta,\vec{p})]\eqend{.}
\end{equation}
Next, we substitute Eq.~(\ref{eq:Fourier_transform_G_Q}) in the expression above. To calculate the limit, we combine Eqs.~(10.4.7) and~(10.7.3) of Ref.~\cite{dlmf} to obtain
\begin{equation}\label{eq:limit_zero_momentum}
 \lim_{\vec{p} \to 0}[H_\mu^{(1)}(-p\eta)H_\mu^{(2)}(-p\eta') - H_\mu^{(1)}(-p\eta')H_\mu^{(2)}(-p\eta)]
 = \frac{2\mathi}{\pi\mu}\frac{(-\eta)^{2\mu} - (-\eta')^{2\mu}}{(\eta\eta')^\mu}\eqend{.}
 \end{equation}
Returning to the computation of $I_{m,\alpha}$, we use the limit above in Eq.~(\ref{eq:I_m_alpha_Fourier}) to obtain
\begin{splitequation}\label{eq:I_m_alpha_eta_0_integral}
 I_{m,\alpha}(\eta, \eta_0) 
 & = -\frac{2}{(n - 2)}\frac{(1 - \epsilon)}{\sqrt{\epsilon}}(-\eta)^{\frac{n - 1}{2} + \nu}(H^\frac{n}{2}a)(\eta)\\
 & \phantom{=} \times \int_{\eta_0}^\eta d\eta'[1 - \epsilon(\eta')]^\frac{n - 2}{2}(\epsilon^{\alpha - \frac{1}{2}}H^{m + \frac{n - 2}{2}}a^n)(\eta')(-\eta')^{\frac{n - 1}{2} - \nu}\eqend{,}
\end{splitequation}
where the parameter $\nu$ is related to the slow-roll parameters as
\begin{equation}\label{eq:nu_definition}
 \nu \equiv \frac{n - 1}{2} + \frac{n - 2}{2}\epsilon + \delta\eqend{.}
\end{equation}

We recall that quantities varying at orders higher than first in the slow-roll parameters are assumed to be constants within the slow-roll approximation. Thus, we can pull the $(1 - \epsilon)$-factor out of the integral, but e.g. must keep the negative powers of $\epsilon$ inside as they can vary up to first order. Then, by using the expressions for $a$, $H$ and $\epsilon$ in terms of the conformal time given in Eqs.~(\ref{eq:slow_roll_ha}), the integral $I_{m,\alpha}$ results in
\begin{equation}\label{eq:I_m_alpha_eta_0}
 I_{m,\alpha}(\eta, \eta_0) = -\frac{2}{(n - 2)\epsilon^{1 - \alpha}} \frac{(H^{m - 1}a)(\eta)}{n - 1 - (m - 1)\epsilon + 2\alpha\delta}\left[1 - \left(\frac{\eta}{\eta_0}\right)^{n - 1 - (m - n)\epsilon + 2\alpha\delta}\right]\eqend{.}
\end{equation}
The term in Eq.~(\ref{eq:I_m_alpha_eta_0}) involving the initial time can be easily expressed in terms of the scale factor $a$. Using Eq.~(\ref{eq:slow_roll_ha}), it can be written as
\begin{equation}
 \left(\frac{\eta}{\eta_0}\right)^{n - 1 - (m - n)\epsilon + 2\alpha\delta} = \left[\frac{a(\eta)}{a(\eta_0)}\right]^{-[n - 1 - (m - 1)\epsilon + 2\alpha\delta]}\eqend{.}
\end{equation}
Although the term above is appreciable at early times, it clearly decreases exponentially during inflation. Considering that the inflationary phase of the early Universe is expected to last for at least around 60 e-folds~\cite{dodelson_cosmology_book}, that term becomes negligible at intermediate and late times and can be dropped, which is equivalent to take the limit $\eta_0 \to -\infty$ in the in-in formalism. Hence, our calculation of the quantum corrections to the expansion rate in slow-roll inflation is accurate only after a large enough number of e-folds has elapsed. The limit~(\ref{eq:I_m_alpha_limit}) in that regime is a good approximation and we are allowed to use
\begin{equation}\label{eq:I_m_alpha}
 I_{m,\alpha}(\eta) = -\frac{2}{(n - 2)\epsilon^{1 - \alpha}} \frac{(H^{m - 1}a)(\eta)}{n - 1 - (m - 1)\epsilon + 2\alpha\delta}\eqend{.}
\end{equation}

Finally, we return to the expression for the contribution from the counter-terms. We substitute Eq.~(\ref{eq:I_m_alpha}) in Eq.~(\ref{eq:CT_integral_I_J}), which yields
\begin{equation}\label{eq:CT_slow_roll}
 \mathi \expect{H^{(1)}(x) S_\mathrm{G,CT}^{(1)}}_0 = \frac{H}{2}\left[\frac{(n - 1)\delta_V}{n - 1 - \epsilon} - \frac{\epsilon(\delta_V - \delta_Z)}{n - 1 - \epsilon + 2\delta}\right]\eqend{.} 
\end{equation}

\subsubsection{The ghost term}
%%%%%%%%%%%%%%%%%%%%%%%%%%%%%%%%%%%%%%%%%%%%%%%%%%%%%%%%%%%%%%%%%%%%%%%%%%
We now consider the contribution coming from the ghost loop, which is given by
\begin{splitequation}\label{eq:GH_integral}
 & \mathi \expect{H^{(1)}(x)S_{\textrm{GH},\textrm{eff}}^{(1)}}_0 
 = -\frac{\mathi}{2(n - 1)a(\eta)}\int d^nx'a^{n - 2}(\eta')\del'^{\nu}F^k_{\phantom{k}k}(x,x')\lim_{y,y' \to x'}\del_{\nu} G^\mathrm{F}_\mathrm{H}(y,y')\\
 &\qquad - \frac{\mathi}{2(n - 1)a(\eta)}\int d^nx'a^{n - 2}(\eta')F^k_{\phantom{k}k}(x,x')\lim_{y,y' \to x'}[\del^2 - (n - 2)(Ha)(y^0)\del_0] G^\mathrm{F}_\mathrm{H}(y,y')\eqend{,}
\end{splitequation}
where $F_{\mu\nu}$ was defined in Eq.~(\ref{eq:F_mu_nu}) and $\del'_\mu$ denotes the partial derivative operator acting on the second argument of the propagator. The operator within the square brackets in the second line of Eq.~(\ref{eq:GH_integral}) is precisely the equation of motion for $G_\mathrm{H}^\mathrm{F}$ and, thus, that limit gives
\begin{equation}
 \lim_{y,y' \to x'}[\del^2 - (n - 2)(Ha)(y^0)\del_0] G^\mathrm{F}_\mathrm{H}(y,y') = \frac{1}{a^{n - 2}(\eta')}\lim_{y,y' \to x'} \delta^{(n)}(y - y')\eqend{.}
\end{equation}
In the dimensional regularisation prescription, however, we have that the coincidence limit of the $\delta$-distribution vanishes---see e.g. Ref.~\cite{leibbrandt_rmp_1975}---, and the second term in Eq.~(\ref{eq:GH_integral}) does not contribute. Moreover, the fact that the state is homogeneous and isotropic allows us to trade $\del'_i$ for $-\del_i$ in the first line of Eq.~(\ref{eq:GH_integral}) and then pull that operator out of the integral. The resulting term will also vanish thanks to the the same symmetries. Hence, Eq.~(\ref{eq:GH_integral}) reduces to
\begin{equation}\label{eq:GH_integral_simplified}
 \mathi \expect{H^{(1)}(x)S_{\textrm{GH},\textrm{eff}}^{(1)}}_0 
 = \frac{\mathi}{2(n - 1)a(\eta)}\int d^nx'a^{n - 2}(\eta')\del_{\eta'}F^k_{\phantom{k}k}(x,x')\lim_{y,y' \to x'}\del_0 G^\mathrm{F}_\mathrm{H}(y,y')\eqend{.}
\end{equation}

The coincidence limit appearing in Eq.~(\ref{eq:GH_integral_simplified}) is given in Eq.~(\ref{eq:coincidence_limit_G_HQ_D_HQ}). Furthermore, from Eq.~(\ref{eq:F_mu_nu}) we have
\begin{equation}
 \del_{\eta'}F^k_{\phantom{k}k}(x,x') = -(n - 1)\left(\epsilon(\eta)\del_\eta\del_{\eta'} - \frac{\laplace}{(Ha)(\eta)}\del_{\eta'}\right)G^\mathrm{c}_\mathrm{Q}(x,x')\eqend{.}
\end{equation}
The contribution of the ghost loop then reduces to
\begin{equation}
 \mathi \expect{H^{(1)}(x)S_{\textrm{GH},\textrm{eff}}^{(1)}}_0 = -\frac{n - 1}{2}\frac{(1 - \epsilon)^n}{a(\eta)}J_{1,\mu,\mu - 1}I_{n,0}(\eta)\eqend{,}
\end{equation}
with the integral $I_{n,0}(\eta)$ as defined in Eqs.~(\ref{eq:I_m_alpha_limit}) and~(\ref{eq:I_m_alpha_integral}). Note that again some of the factors involving $\epsilon$ varying at order higher than one in the slow-roll approximation have been already pulled out of the integral. Finally, by using Eqs.~(\ref{eq:I_m_alpha}) and~(\ref{eq:J_integral_values}), we obtain
\begin{equation}
 \mathi \expect{H^{(1)}(x)S_{\textrm{GH},\textrm{eff}}^{(1)}}_0 = -H^{n - 1}D_\mathrm{GH}(n,\epsilon,\delta)\eqend{,}
\end{equation}
with
\begin{equation}\label{eq:D_GH}
 D_\mathrm{GH}(n,\epsilon,\delta) = \frac{nA^{(n)}_\mu}{2}(1 - \epsilon)^{n - 1}(2 + \epsilon)\eqend{.}
\end{equation}

\subsubsection{Three-gravitons interaction: the \texorpdfstring{$V$}{V}-tensor term}
%%%%%%%%%%%%%%%%%%%%%%%%%%%%%%%%%%%%%%%%%%%%%%%%%%%%%%%%%%%%%%%%%%%%%%%%%%
The contribution from the three-gravitons interaction term involving the tensor $V^{\alpha\beta\mu\nu\sigma\rho}$ is given by
\begin{splitequation}\label{eq:V_contribution}
 & \mathi \expect{ H^{(1)}(x)S^{(1)}_{\textrm{G}, V}}_0\\
 & \qquad = -\frac{\mathi}{8}\frac{(n - 2) V^{\alpha\beta\mu\nu\sigma\rho}}{(n - 1)a(\eta)} \int d^nx'(Ha^{n - 1})(\eta')\left[F_{\alpha\beta}(x,x')\lim_{y,y' \to x'}\del_\rho G^\mathrm{F}_{0\sigma\mu\nu}(y,y')\right.\\
 & \qquad \phantom{=} \quad \left. + F_{0\sigma}(x,x')\lim_{y,y' \to x'}\del_\rho G^\mathrm{F}_{\mu\nu\alpha\beta}(y,y') + \del'_\rho F_{\mu\nu}(x,x') \lim_{y,y' \to x'} G^\mathrm{F}_{\alpha\beta0\sigma}(y,y')\right]\eqend{,}
\end{splitequation}
and we remind the reader that $F_{\mu\nu}$ was defined in Eq.~(\ref{eq:F_mu_nu}). Besides the components of $F_{\mu\nu}$ already given in Eqs.~(\ref{eq:F_00_slow_roll}) and (\ref{eq:F_ij_slow_roll}), here we also need
\begin{equation}\label{eq:F_0i_slow_roll}
 F_{0i}(x,x') = \del_i\left[\frac{\epsilon(\eta)\epsilon(\eta')}{2}D^\mathrm{c}_\mathrm{Q}(x,x') - \frac{(\epsilon Ha)(\eta)\del_\eta + (\epsilon Ha)(\eta')\del_{\eta'} - \laplace}{2(Ha)(\eta)(Ha)(\eta')}G^\mathrm{c}_\mathrm{Q}(x,x')\right]\eqend{.}
\end{equation}
The terms in Eq.~(\ref{eq:V_contribution}) involving total spatial derivatives at the observation point $x$ vanish when integrated, and we are left with
\begin{splitequation}\label{eq:V_contribution_simplified}
 & \mathi \expect{H^{(1)}(x)S^{(1)}_{\textrm{G}, V}}_0\\
 & = -\frac{\mathi}{8}\frac{n - 2}{(n - 1)a(\eta)} \int d^nx'(Ha^{n - 1})(\eta')\left[V^{ij\mu\nu\sigma\rho}F_{ij}(x,x')\lim_{y,y' \to x'}\del_\rho G^\mathrm{F}_{0\sigma\mu\nu}(y,y')\right.\\
 & \phantom{=} \qquad\qquad\qquad\qquad \left. V^{\alpha\beta ij\sigma\rho}\del_{\eta'} F_{ij}(x,x') \lim_{y,y' \to x'} G^\mathrm{F}_{\alpha\beta0\sigma}(y,y')\right]\eqend{.}
\end{splitequation}

The details of the computation of the coincidence limits appearing in the expression above can be found in \ref{apdx:coincidence_limit_three_gravitons}. The result is
\begin{splitequation}\label{eq:V_contribution_J}
 & \mathi \expect{H^{(1)}(x)S^{(1)}_{\textrm{G}, V}}_0\\
 & = \frac{(1 - \epsilon)^n\epsilon}{8(n - 1)a(\eta)}\bigg\{-[(2n^2 - 7n + 7)\epsilon + 8(n - 1)\delta]J_{2,\nu - 1,\nu - 1} + (n - 1)(1 - \epsilon)J_{3,\nu,\nu - 1}\\ 
 & \phantom{=} - (n - 1)[(n - 1)(2 - \epsilon) - 2\delta]J_{2,\nu,\nu} + \frac{(n - 1)[4(n - 3)(n - 1) - (2n^2 - 7n - 3)\epsilon]}{1 - \epsilon}\\ 
 & \phantom{=} \times J_{1,\nu,\nu - 1} - (n - 1)^2\epsilon^2 J_{0,\nu - 1,\nu - 1}\bigg\}I_{n,-1}(\eta)\\
 & \phantom{=} + \frac{(1 - \epsilon)^{n - 1} \epsilon^2}{8a(\eta)}\bigg\{\frac{(n - 5)(n - 2)}{n - 1}(J_{2,\mu,\mu} - J_{2,\mu - 1, \mu - 1}) + (2n^3 - 8n^2 + 11n - 9)J_{1,\mu,\mu - 1}\\ 
 & \phantom{=} - (n - 1)[(n - 3)(n^2 - 3n + 3) + (n - 2)^2\epsilon]J_{0,\mu - 1,\mu - 1}\bigg\}I_{n,-1}(\eta)\eqend{,}
\end{splitequation}
where the integral $I_{n,-1}$ was defined in Eqs.~(\ref{eq:I_m_alpha_limit}) and~(\ref{eq:I_m_alpha_integral}). We again cast the expression above as
\begin{equation}
 \mathi \expect{H^{(1)}(x)S^{(1)}_{\textrm{G}, V}}_0 = -H^{n - 1}D_{\mathrm{G},V}(n,\epsilon,\delta)\eqend{,}
\end{equation}
only to obtain
\begin{splitequation}
 D_{\mathrm{G},V}(n,\epsilon,\delta)
 & = \frac{(1 - \epsilon)^{n - 1}A^{(n)}_\mu}{8(n - 2)[(n - 1)(1 - \epsilon) - 2\delta]}\left[2n(11 - 14n + 4n^3 - 5n^4) - n(6 - 23n \right.\\
 & \phantom{=} \qquad \left. + 28n^2 - 13n^3 + 2n^4)\epsilon\right]\\
 & \phantom{=} + \frac{(1 - \epsilon)^{n - 1}A^{(n)}_\nu}{8(n - 2)[(n - 1)(1 - \epsilon) - 2\delta]\epsilon}\left[2(2 + 15n - 30n^2 + 15n^3 - 2n^4)\right.\\
 & \phantom{=} \qquad \left. + (26 + 3n - 16n^2 + 10n^3 - 3n^4)\epsilon - 2(13 - 8n - 8n^2 + 3n^3)\delta\right]\eqend{.}
\end{splitequation}

\subsubsection{Three-gravitons interaction: the \texorpdfstring{$U$}{U}-tensor term}
%%%%%%%%%%%%%%%%%%%%%%%%%%%%%%%%%%%%%%%%%%%%%%%%%%%%%%%%%%%%%%%%%%%%%%%%%%
The interaction in Eq.~(\ref{eq:S_G_U}) contributes with the term
\begin{splitequation}\label{eq:U_contribution}
 & \mathi \expect{H^{(1)}(x)S^{(1)}_{\textrm{G}, U}}_0\\
 & \qquad = -\frac{\mathi}{16}\frac{U^{\alpha\beta\gamma\delta\mu\nu\rho\sigma}}{(n - 1)a(\eta)}\int d^nx'a^{n - 2}(\eta')\left[F_{\gamma\delta}(x,x')\lim_{y,y' \to x'} \del_\alpha\del_\beta'G^\mathrm{F}_{\mu\nu\rho\sigma}(y,y')\right.\\
 & \qquad \phantom{=} \left. + \del_\alpha'F_{\mu\nu}(x,x')\lim_{y,y' \to x'} \del_\beta G^\mathrm{F}_{\rho\sigma\gamma\delta}(y,y') + \del_\beta'F_{\rho\sigma}(x,x')\lim_{y,y' \to x'} \del_\alpha G^\mathrm{F}_{\mu\nu\gamma\delta}(y,y')\right]\eqend{,}
\end{splitequation}
with $F_{\mu\nu}$ given by Eq.~(\ref{eq:F_mu_nu}). We again refer the reader to \ref{apdx:coincidence_limit_three_gravitons} for the details of the computation of the coincidence limits appearing above. The final result is
\begin{splitequation}\label{eq:U_contribution_J}
 \mathi \expect{H^{(1)}(x)S^{(1)}_{\textrm{G}, U}}_0
 & = \frac{[(1 - \epsilon)H]^{n - 1}\epsilon^2}{16(n - 1)a(\eta)}\left\{2(n - 1)^2(-3 + 12n - 10n^2 + 2n^3) J_{1,\mu,\mu - 1}
 \right.\\
& \phantom{=} + (n - 1)(-106 + 100n - 35n^2 + 5n^3) J_{2,\mu - 1,\mu - 1} - (26 - 122n \\
& \phantom{=} \left. + 109n^2 - 38n^3 + 5n^4) J_{2,\mu,\mu} + 8(n - 5)(n - 2)J_{3,\mu,\mu - 1}\right\}I_{n,-1}(\eta)\\
& \phantom{=} - \frac{[(1 - \epsilon)H]^{n - 1}\epsilon}{16(n - 1)a(\eta)}\left\{2\left[2(n - 5)(n - 2)(n - 1)^2 - (n - 1)^2(27 - 16n\right.\right.\\ 
& \phantom{=} \left.+ 2n^2)\epsilon - 4(n - 5)(n - 2)(n - 1)\delta\right]J_{1,\nu,\nu - 1} + 2\left[(n - 2)(n - 1)^{\phantom{1}}\right.\\ 
& \phantom{=} \left. \times (3n - 7) + (11 - 36n + 24n^2 - 5n^3)\epsilon - 4(n - 2)(n - 1)\delta\right]J_{2,\nu - 1,\nu - 1}\\
& \phantom{=} - 2\left[2(-17 + 17n - 7n^2 + n^3) + (31 - 33n + 17n^2 - 3n^3)\epsilon\right.\\ 
& \phantom{=} \left.- 2(n - 3)(n + 1)^{\phantom{1}}\!\!\delta\right]J_{2,\nu,\nu} + 2\left[31 - 15n + 2n^2 - (63 - 32n + 5n^2)\epsilon\right.\\
& \!\left.\phantom{1^2} \left. \!\!\!\!\!\!\!\phantom{1^1}- 4(n - 2)\delta\right]J_{3,\nu,\nu - 1} - 2(n - 3)(1 - 3\epsilon)(J_{4,\nu - 1,\nu - 1} - J_{4,\nu,\nu})\right\}I_{n,-1}(\eta)\eqend{.}
\end{splitequation}
Finally, we use Eqs.~(\ref{eq:I_m_alpha}) and~(\ref{eq:J_integral_values}) to cast the expression above in the form
\begin{equation}
 \mathi \expect{H^{(1)}(x)S^{(1)}_{\textrm{G}, U}}_0 = -H^{n - 1}D_{\mathrm{G},U}(n,\epsilon,\delta)\eqend{,}
\end{equation}
with
\begin{splitequation}
 D_{\mathrm{G},U}(n,\epsilon,\delta)
 & = \frac{(1 - \epsilon)^{n - 1}A^{(n)}_\mu}{8(n - 2)^2[(n - 1)(1 - \epsilon) - 2\delta]}(n - 1)n(n - 2)^2(36 - 11n + n^2) \\
 & \phantom{=} - \frac{(1 - \epsilon)^{n - 1}A^{(n)}_\nu}{8(n - 2)^2(n + 2)[(n - 1)(1 - \epsilon) - 2\delta]\epsilon}\left[4(n - 2)(n - 1)(11 - 13n \phantom{1^2}\right.\\ 
 & \phantom{=} - 5n^2 + 4n^3) - (n - 2)(36 + 23n + 14n^2 - 23n^3 - 17n^4 + 9n^5)\epsilon\\ 
 & \phantom{=} \left. - 2(120 - 57n - 38n^2 + 49n^3 - 25n^4 + 5n^5)\delta\right]\eqend{.}
\end{splitequation}

\subsection{Renormalisation}
%%%%%%%%%%%%%%%%%%%%%%%%%%%%%%%%%%%%%%%%%%%%%%%%%%%%%%%%%%%%%%%%%%%%%%%%%%%%%%%%%%%%%%%%%%%%%%%%%%%%%%%%%%%%%%%%%%%%%%%%%%%%%%%%%%%%%%%%%%%%%%%%%%%%%

Here we are dealing with a composite operator whose divergences cannot be fully absorbed in the renormalisation of the $N$-point functions of the basic fields alone. It is known~\cite{itzykson_and_zuber_qft_book, bonneau_scholarpedia_2009, hollands_wald_pr_2015} that, apart from the usual counter-terms in the bare Lagrangian, one also needs counter-terms coming from all the operators that can mix with $\mathcal{H}$. They are all the operators with the same quantum numbers (spin, charges, etc.) as and having equal or lower dimension than $\mathcal{H}$, in general. There would be just a finite number of such operators, were we analysing a local observable, but for non-local observables like $\mathcal{H}$ the combinations are endless and no general framework is available in the literature to determine them. The only example of renormalisation of a non-local operator that is well understood is the Wilson loop in non-Abelian gauge theories~\cite{korchemsky_radyushkin_npb_1987}. 

Inspired by the Wilson-loop case and their results in de~Sitter space-time, Miao~{\it et~al}~\cite{miao_tsamis_woodard_prd_2017} have conjectured that the operators $\mathcal{R}\mathcal{H}$ and $\mathcal{H}^3$, where $\mathcal{R}\equiv R[x(\tilde{X})]$ corresponds to the gauge-invariant Ricci scalar defined as in Eq.~(\ref{eq:scalar_observable}), should be enough to renormalise the invariant expansion rate on general FLRW background space-times, at least at one-loop order. Shortly after, Fr{\"o}b~\cite{froeb_cqg_2019} showed that those operators and the operator $\mathcal{H}$ itself are enough to renormalise the invariant expansion rate at one-loop order in single-field inflation and with constant $\epsilon$.

We start this calculation by recalling that in the slow-roll approximation only quantities that vary slowly in time, i.e.\ whose time derivative is second order or higher in the slow-roll parameters, are taken as constants. That is precisely the case of the slow-roll parameters $\epsilon$ and $\delta$. Thus, after expanding the counter-terms contribution of Eq.~(\ref{eq:CT_slow_roll}) up to first order in the slow-roll parameters, we see that we can take $\delta_Z = 0$, just as in the constant-$\epsilon$ case. This reduces that equation to
\begin{equation}
i\left\langle H^{(1)}(x)S^{(1)}_\mathrm{G,CT}\right\rangle = \kappa^2 \frac{H}{2}\delta_V\eqend{.} 
\end{equation}
Moreover, it is convenient to define
\begin{equations}[eq:D_total]
 D(n, \epsilon, \delta) & \equiv D_1(n,\epsilon,\delta) + D_2(n,\epsilon,\delta)\eqend{,}\\
 D_1(n,\epsilon,\delta) & \equiv D_\mathrm{GH}(n,\epsilon,\delta) + D_{\mathrm{G},V}(n,\epsilon,\delta) + D_{\mathrm{G},U}(n,\epsilon,\delta)\eqend{.}
\end{equations}
A simple computation then gives
\begin{equation}\label{eq:D_total_expression}
 D(n, \epsilon, \delta) = \frac{1}{n - 4}\frac{1}{768\pi^2}\left[63\left(1 + \frac{\delta}{\epsilon}\right) - 4539\epsilon - 103\delta + 76\frac{\delta^2}{\epsilon}\right] + \bigo{(n - 4)^0}\eqend{.}
\end{equation}
It is not difficult to conclude from Eq.~(\ref{eq:slow_roll_ha}) that the term $\delta/\epsilon$ cannot be taken as a constant, a priori. This is because its time derivative is only first order in the slow-roll approximation. All the other terms in Eq.~(\ref{eq:D_total_expression}), however, vary slowly in time and, thus, can be well approximated by constants. Hence, in principle, we must find another operator that mix with $\mathcal{H}$ and its counter-terms is able to absorb the divergence of the term involving 
$\delta/\epsilon$.

In any case, let us assume for a moment that $\delta/\epsilon$ is a constant. By making this assumption we are introducing in the expression for $D(n,\epsilon,\delta)$ an error at first order in the slow-roll approximation. The renormalisation procedure in this case becomes similar to the constant-$\epsilon$ one, and we can use the same counter-terms. The renormalised result then reads
\begin{splitequation}
 \langle\mathcal{H}_\mathrm{ren}(x)\rangle 
 & = \lim_{n \to 4}\left[H - \kappa^2H^{n - 1}D(n,\epsilon,\delta) + \kappa^2 \frac{H}{2}\delta_V\right.\\ 
 & \left. \phantom{\frac{1}{1}}\qquad + \kappa^2\mu^{n - 4}(\mathcal{H}^3)_0\alpha - \kappa^2\mu^{n - 2}\mathcal{H}_0\beta\right]\eqend{,}
\end{splitequation}
where $\mu$ now denotes a renormalisation scale with dimension of mass. We have only considered the counter-term coming from $\mathcal{H}^3$ as we cannot distinguish it from $\mathcal{R}\mathcal{H}$ at this loop order. We choose the renormalisation scale to be $H_0$, $\delta_V$ to cancel the divergences coming from the graviton one-point function at the initial time $\eta_0$~\cite{miao_tsamis_woodard_prd_2017, tsamis_woodard_annp_2006} and $\alpha$ and $\beta$ to cancel the divergences in $D(n,\epsilon,\delta)$ and $D_1(n,\epsilon,\delta)$ when $\eta \neq \eta_0$. Thus, since the background values of the mixing operators are $(\mathcal{H}^3)_0 = H^3$ and $\mathcal{H}_0 = H$, these choices amount to take
\begin{equation}
 \delta_V = 2H_0^{n - 2}D_1(n,\epsilon,\delta)\eqend{,}\;\;\; \alpha = D(n,\epsilon,\delta)\eqend{,}\;\;\; \beta = D_1(n,\epsilon,\delta)\eqend{.}
\end{equation}
We then obtain
\begin{equation}\label{eq:invariant_H_renormalised_final_slow_roll_1}
 \langle\mathcal{H}_\mathrm{ren}(x)\rangle = H + \kappa^2\epsilon H^3\ln a\lim_{n \to 4}[(n - 4)D(n,\epsilon,\delta)]\eqend{.}
\end{equation}
This result is correct up to first order in the slow-roll parameters since the error in assuming $\delta/\epsilon$ constant in the expression for $D(n,\epsilon,\delta)$ is pushed to the next order due to the extra $\epsilon$ factor.

We can also keep $\delta/\epsilon$ as a function of time and introduce a new counter-term to absorb the corresponding divergence. Then, we need an extra counter-term coming from an operator that is proportional to $H^3/\epsilon$ on the background. As expected, the list of such operators is endless. It is clear, however, that none of these operators can be written as polynomials of derivatives of the metric alone. We then choose the operator
\begin{equation}\label{eq:other_mixing_operator}
 \frac{\mathcal{H}^5}{\sqrt{-\tilde{\nabla}^\sigma\mathcal{H}\tilde{\nabla}_\sigma\mathcal{H}}}\eqend{.}
\end{equation}
This operator might look as an unusual choice at first sight, but we note e.g.\ that the operator measuring the local expansion rate was defined by a similar formula---see Eq.~(\ref{eq:local_Hubble_rate_definition}).\footnote{Operators such as $\mathcal{H}$ or the one in Eq.~(\ref{eq:other_mixing_operator}) are defined only perturbatively, i.e. in terms of a power series in the basic fields $\phi^{(1)}$ and $h_{\mu\nu}$.} On the background, the operator we suggest reads
\begin{equation}\label{eq:new_mixing_operator}
 \left(\frac{\mathcal{H}^5}{\sqrt{-\tilde{\nabla}^\sigma\mathcal{H}\tilde{\nabla}_\sigma\mathcal{H}}}\right)_0 = \frac{H^3}{\epsilon}\eqend{.}
\end{equation}
The expectation value of the renormalised $\mathcal{H}$ then is
\begin{splitequation}
 \langle\mathcal{H}_\mathrm{ren}(x)\rangle 
 & = \lim_{n \to 4}\left[H - \kappa^2H^{n - 1}D(n,\epsilon,\delta) + \kappa^2 \frac{H}{2}\delta_V + \kappa^2\mu^{n - 4}(\mathcal{H}^3)_0\alpha \phantom{\left(\frac{I^1}{\sqrt{-\tilde{I}^i}}\right)_I}\right. \\ 
 & \left.\qquad\qquad - \kappa^2\mu^{n - 2}\mathcal{H}_0\beta + \kappa^2\mu^{n - 4}\left(\frac{\mathcal{H}^5}{\sqrt{-\tilde{\nabla}^\sigma\mathcal{H}\tilde{\nabla}_\sigma\mathcal{H}}}\right)_0\gamma\right]\eqend{.}
\end{splitequation}
In order to simplify the notation, let us write
\begin{equation}
 D(n,\epsilon,\delta) = \frac{1}{n - 4}\left(a + \frac{b}{\epsilon}\right) + \bigo{(n - 4)^0}\eqend{,}
\end{equation}
with $a$ and $b$ constants. We choose the same renormalisation conditions as above, which now correspond to
\begin{equation}
 \delta_V = 2H_0^{n - 2}D_1(n,\epsilon,\delta)|_{\eta = \eta_0}\eqend{,}\;\;\; \alpha = \frac{a}{n - 4}\eqend{,}\;\;\; \beta = \frac{2}{n - 4}D_1(n,\epsilon,\delta)|_{\eta = \eta_0} \eqend{,}\;\;\; \gamma = \frac{b}{n - 4}\eqend{.}
\end{equation}
Then,
\begin{splitequation}
 \langle\mathcal{H}_\mathrm{ren}(x)\rangle 
 & = H - \kappa^2\lim_{n \to 4}H^{n - 1}\left[D(n,\epsilon,\delta) - \left(\frac{H_0}{H}\right)^{n - 2}D_1(n,\epsilon,\delta)|_{\eta = \eta_0} - \left(\frac{H_0}{H}\right)^{n - 4}\frac{a}{n - 4} \right. \\ 
 & \phantom{=} \qquad\qquad\qquad\qquad\left. + \left(\frac{H_0}{H}\right)^{n - 2}D_1(n,\epsilon,\delta)|_{\eta = \eta_0} - \left(\frac{H_0}{H}\right)^{n - 4}\frac{b}{n - 4}\frac{1}{\epsilon}\right]\\
 & = H + \kappa^2\epsilon H^2\ln a\lim_{n \to 4}[(n - 4)D(n,\epsilon,\delta)]\eqend{,}
\end{splitequation}
which clearly agrees with the results obtained by treating all terms in $D(n,\epsilon,\delta)$ as constants, as it should.

\subsection{Results}
%%%%%%%%%%%%%%%%%%%%%%%%%%%%%%%%%%%%%%%%%%%%%%%%%%%%%%%%%%%%%%%%%%%%%%%%%%%%%%%%%%%%%%%%%%%%%%%%%%%%%%%%%%%%%%%%%%%%%%%%%%%%%%%%%%%%%%%%%%%%%%%%%%%%%

We use Eq.~(\ref{eq:D_total_expression}) in Eq.~(\ref{eq:invariant_H_renormalised_final_slow_roll_1}) to obtain
\begin{equation}\label{eq:invariant_Hubble_rate_slow_roll}
 \langle\mathcal{H}_\mathrm{ren}(x)\rangle = H\left[1 + \frac{63}{768\pi^2}\kappa^2(\epsilon + \delta) H^2\ln a\right]\eqend{.}
\end{equation}
Thus, the invariant expansion rate receives a finite quantum correction at one-loop order that produces a secular effect, i.e. produces terms that grow in time. It follows from the cumulative effect of gravitons being copiously produced by the background expansion~\cite{ford_prd_1985, tsamis_woodard_plb_1993}. As these quanta pile up, the quantum fluctuations of the metric experience a slow growth, and this is reflected in the expectation values of the observables. For sufficiently rapid background expansion, this effect can become strong enough to cause the perturbative treatment to break down. 

We can estimate how long it takes to the one-loop correction to become comparable to the background using the current experimental constraints~\cite{planck_2018c}. The upper bound for the energy scale of inflation is of order $10^{16}\, \mathrm{GeV}$, thus we have $\kappa^2H^2 < (10^{16}\, \mathrm{GeV}/E_\mathrm{P})^4$, where $E_\mathrm{P}$ is the Planck energy, while the first and second slow-roll parameter we can estimate as $|\epsilon + \delta|\approx 10^{-3}$. Thus, the loop correction in Eq.~(\ref{eq:invariant_Hubble_rate_slow_roll}) becomes order one after no less than a number of e-folds of order $10^{17}$. Although the one-loop correction needs a huge number of e-folds to become important, this is not rule out as there is no upper bound on the duration of inflation. It is possible, however, that this number is brought down by the higher-loop correction. These corrections will involve higher powers of $\ln a$, thus as the perturbative approach breaks down all the other loop orders become important, making the actual number of e-folds to be smaller than the crude one-loop estimation.

In any case, when these corrections become important one needs to employ some kind of resummation method to determine the late-time behaviour of the quantum backreaction. In the slow-roll case, for example, we can try to estimate how this non-perturbative regime looks like by quantum correcting the first slow-roll parameter~\cite{froeb_cqg_2019}. Hence, let us write the term multiplying $H$ in Eq.~(\ref{eq:invariant_Hubble_rate_slow_roll}) as
\begin{equation}
 1 + \frac{63}{768\pi^2}\kappa^2(\epsilon + \delta) H^2\ln a + \bigo{\epsilon^2, \delta^2}= a^{\frac{63}{768\pi^2}\kappa^2 H_0^2(\epsilon + \delta)} + \bigo{\epsilon^2, \delta^2}\eqend{.}
\end{equation}
Then, going back to Eq.~(\ref{eq:invariant_Hubble_rate_slow_roll}) and using Eq.~(\ref{eq:slow_roll_ha}), we have
\begin{equation}
 \langle\mathcal{H}_\mathrm{ren}(x)\rangle = H(\hat{\epsilon}) + \bigo{\epsilon^2, \delta^2}\eqend{,}
\end{equation}
with the quantum-corrected deceleration parameter
\begin{equation}\label{eq:quantum_corrected_epsilon}
 \hat{\epsilon} = \epsilon - \frac{63}{768\pi^2}\kappa^2 H_0^2 (\epsilon + \delta)\eqend{.}
\end{equation}
In the case the second slow-roll parameter $\delta = 0$ the quantum correction shifts $\epsilon$ towards the de~Sitter space-time, where $\epsilon = 0$~\cite{froeb_cqg_2019}. For finite $\delta$, however, the backreaction might move us towards or away from the de~Sitter space-time, depending on the magnitude and sign of the second slow-roll parameter. This shift, however, is quite small. Using again the experimental constraints on the cosmological parameters, we have that the background first slow-roll parameter is shifted by a number smaller than $10^{-17}$.

%%%%%%%%%%%%%%%%%%%%%%%%%%%%%%%%%%%%%%%%%%%%%%%%%%%%%%%%%%%%%%%%%%%%%%%%%%%%%%%%%%%%%%%%%%%%%%%%%%%%%%%%%%%%%%%%%%%%%%%%%%%%%%%%%%%%%%%%%%%%%%%%%%%%%
\section{Conclusions}                                                                                                                               %
\label{sec:conclusions}                                                                                                                             %
%%%%%%%%%%%%%%%%%%%%%%%%%%%%%%%%%%%%%%%%%%%%%%%%%%%%%%%%%%%%%%%%%%%%%%%%%%%%%%%%%%%%%%%%%%%%%%%%%%%%%%%%%%%%%%%%%%%%%%%%%%%%%%%%%%%%%%%%%%%%%%%%%%%%%

We have calculated the graviton one-loop correction to the expectation value to an observable that measures the local expansion rate in slow-roll inflation. The operator corresponding to this observable is of the relational type and fully gauge invariant. Similar calculations were recently performed in de~Sitter space-time~\cite{miao_tsamis_woodard_prd_2017} and in single-field inflation with constant $\epsilon$~\cite{froeb_cqg_2019}. 

The de~Sitter calculation has confirm the expectation that there is no one-loop backreaction on the background expansion rate. It is based on the observable proposed in Ref.~\cite{tsamis_woodard_prd_2013}, which requires the use of a non-local time coordinate as in that case there is no scalar degree of freedom that can serve as a clock. In Sec.~\ref{sec:gauge_invariant_observables} we have shown how that proposal fits within the framework of Refs.~\cite{brunetti_et_al_jhep_2016, froeb_cqg_2018, froeb_lima_cqg_2018} by considering a simple generalisation of the equation defining the configuration-dependent coordinates $\tilde{X}^{(\alpha)}[\tilde{g}]$, see Eq.~(\ref{eq:general_coordinates}). 

The constant-$\epsilon$ calculation is based on the same observable we have employed here to compute the one-loop correction of the expansion rate. However, there are some differences between that calculation and the one we have presented in Sec.~\ref{sec:one_loop_correction} that are important to highlight. Firstly, the in-in formalism cannot be employed straightaway in the case of the slow-roll approximation. For the in-in formalism relies on the assumption that the initial state can be given asymptotically far in the past, while the slow-roll approximation is valid only for a finite time interval. We have shown that although the contribution from placing the initial state at a certain finite initial time is appreciable at early times, it can be neglected at late times and is equivalent to send the initial time to the far past. Secondly, although the renormalisation in the slow-roll case can be performed just as in the constant-$\epsilon$ space-times, a calculation that keeps track of the slow-roll approximation is also possible. The latter requires an extra counter-term coming from e.g.\ the operator in Eq.~(\ref{eq:other_mixing_operator}), although it is not clear if the precise form of that operator is an artefact of the slow-roll approximation. In any case, this calculation provides a partially negative answer to the question raised in Ref.~\cite{froeb_cqg_2019} of whether the slow-roll case could differentiate between the counter-terms coming from $\mathcal{R}\mathcal{H}$ and $\mathcal{H}^3$ and/or unveil counter-terms beyond the ones conjectured in Ref.~\cite{miao_tsamis_woodard_prd_2017} at one-loop order. Moreover, the quantum-corrected first slow-roll parameter, Eq.~(\ref{eq:quantum_corrected_epsilon}), shows that the graviton backreaction in slow-roll inflation can either accelerate or decelerate the background expansion, which is qualitatively different from the result obtained in the constant-$\epsilon$ case.

The one-loop correction we have obtained grows in time with the logarithm of the scale factor. Thus, at some point the quantum-gravitational backreaction is not the one suggested by the one-loop result we have obtained, and some resummation method is required to predict the backreaction at late times. Using the current experimental data~\cite{planck_2018c}, we have estimated that it takes at least $10^{17}$ e-folds to the one-loop result to become order $1$. This is certainly a huge number. Nevertheless, currently there is no restriction on the maximum duration of the inflationary era. It is important to notice, however, that as the system moves away from the perturbative regime, all the higher-loop corrections contribute very strongly sooner than the one-loop correction.

One can wonder whether this secular effect is an artefact produced by the way perturbation theory is set up, with the dressed interacting state built by switching on the interaction adiabatically from an initial free state lying far in the past. If secular terms were to develop by somehow being already present in the initial state and not by effect of the background dynamics, then it should be possible to absorb them into a redefinition of the initial state. It would be interesting to check whether this is actually the case for the problem at hand. However, the experience with loop corrections in the scalar case~\cite{onemli_woodard_prd_2004, kahya_onemli_woodard_prd_2010} suggests that, even though powers of the scale factor can be absorbed this way, its logarithm cannot. This is in line with other cases where logarithmic terms are important and renormalisation-group techniques to resum them are required~\cite{itzykson_and_zuber_qft_book}.

Another issue is how the one makes contact between the loop correction we have computed here and the observed value for the cosmological parameters. The observable we have considered represents a possible measurement of the local expansion rate during the inflationary era. A measurement of the expansion rate at that time would measure the full $\left\langle \mathcal{H}_\mathrm{ren}(x)\right\rangle$. To disentangle the background value of $\mathcal{H}$ from its quantum corrections one would have to known what was the mechanism driving primordial inflation. Without such mechanism, one can speculate on what would be the role played low-energy quantum gravity during the inflationary era. For example, it has been long advocated~\cite{woodard_ijmpd_2014} that the backreaction of the gravitons on the space-time expansion rate produced during inflation could provide a natural mechanism to a graceful exit from the inflationary era, avoiding the need of fine tuning the potential. The gauge-invariant measure of the one-loop backreaction we have calculated here seems to indicate that indeed quantum corrections could help to decelerate the space-time expansion, depending on the values of $\epsilon$ and $\delta$. This one-loop effect alone, however, does not seem to be enough as it produces a small deviation from the quasi-de~Sitter background, but a two-loop calculation could show more promising results~\cite{tsamis_woodard_npb_1996, miao_tsamis_woodard_prd_2017}. 

%To make a more direct contact between the CMB power spectrum and the observable $\mathcal{H}$, the natural way would be to look at two-point correlation function of the observable. Since $\mathcal{H}$ fluctuates, different points should experience different expansion rates, which could have an impact on the experimental data. Hence, it would be interesting to compute loop corrections to $\left\langle \mathcal{H}(x)\mathcal{H}(y)\right\rangle$, and we hope to report on this in the future.

\ack

The author thanks Atsushi Higuchi and Markus Fr{\"o}b for discussions, and Markus Fr{\"o}b for helping him to check and correct the expressions in the constant-$\epsilon$ case. This work was supported by the Grant No.~RPG-2018-400, ``Euclidean and in-in formalisms in static space-times with Killing horizons'', from the Leverhulme Trust.

\appendix

%%%%%%%%%%%%%%%%%%%%%%%%%%%%%%%%%%%%%%%%%%%%%%%%%%%%%%%%%%%%%%%%%%%%%%%%%%%%%%%%%%%%%%%%%%%%%%%%%%%%%%%%%%%%%%%%%%%%%%%%%%%%%%%%%%%%%%%%%%%%%%%%%%%%%
\section{Free propagators}                                                                                                                          %
\label{apdx:free_propagators}                                                                                                                       %
%%%%%%%%%%%%%%%%%%%%%%%%%%%%%%%%%%%%%%%%%%%%%%%%%%%%%%%%%%%%%%%%%%%%%%%%%%%%%%%%%%%%%%%%%%%%%%%%%%%%%%%%%%%%%%%%%%%%%%%%%%%%%%%%%%%%%%%%%%%%%%%%%%%%%

The linear theory based on the exact gauge~(\ref{eq:gauge_conditions}) was studied in Ref.~\cite{froeb_lima_cqg_2018}, and the corresponding free propagators in a spatially homogeneous and isotropic space-time were derived. Here we quote the expressions from this reference which are pertinent to our calculation. 

The expressions for the Feynman propagator components for general $\epsilon$ and $\delta$ are
\begin{equations}[eq:graviton_propagator]
 G^\textrm{F}_{0000}(x,x') &= \frac{1}{(Ha)(\eta) (Ha)(\eta')} \partial_\eta \partial_{\eta'} G^\text{F}_\text{Q}(x,x') \eqend{,} \\
 G^\textrm{F}_{000k}(x,x') &= \frac{\epsilon(\eta')}{2 (Ha)(\eta)} \partial_k D^\text{F}_\text{Q}(x,x') - \frac{1}{2 (Ha)(\eta) (Ha)(\eta')} \partial_\eta \partial_k G^\text{F}_\text{Q}(x,x') \eqend{,} \\
 G^\textrm{F}_{00kl}(x,x') &= - \delta_{kl} \frac{1}{(Ha)(\eta)} \partial_\eta G^\text{F}_\text{Q}(x,x') \eqend{,} \\
\begin{split}
 G^\textrm{F}_{0i0k}(x,x') &= \Pi_{ik} \left[ D^\text{F}_\text{H}(x,x') + D^\text{F}_2(x,x') \right] + \frac{\partial_i \partial_k}{\laplace} \bigg[ \frac{n-1}{2 (n-2)} D^\text{F}_\text{H}(x,x') \\
&\qquad\quad+ D^\text{F}_2(x,x') + \frac{(\epsilon H a)(\eta) \partial_\eta + (\epsilon H a)(\eta') \partial_{\eta'} - \laplace}{4 (H a)(\eta) (H a)(\eta')} G^\text{F}_\text{Q}(x,x') \\
&\qquad\quad- \frac{\epsilon(\eta) \epsilon(\eta')}{4} D^\text{F}_\text{Q}(x,x') \bigg] \eqend{,}
\end{split} \\
\begin{split}
 G^\textrm{F}_{0ikl}(x,x') &= - 2 \frac{\delta_{i(k} \partial_{l)}}{\laplace} \partial_\eta G^\text{F}_2(x,x') \\
&\quad- \delta_{kl} \frac{\partial_i}{\laplace} \left[ \frac{1}{n-2} \partial_\eta G^\text{F}_\text{H}(x,x') - \left[ \frac{\epsilon(\eta)}{2} \partial_\eta - \frac{\laplace}{2 (H a)(\eta)} \right] G^\text{F}_\text{Q}(x,x') \right] \eqend{,}
\end{split} \\
\begin{split}
 G^\textrm{F}_{ijkl}(x,x') &= \left( 2 \delta_{i(k} \delta_{l)j} - \frac{2}{n-2} \delta_{ij} \delta_{kl} \right) G^\text{F}_\text{H}(x,x') + \delta_{ij} \delta_{kl} G^\text{F}_\text{Q}(x,x') \\
&\quad- 4 \frac{\partial_{(i} \delta_{j)(k} \partial_{l)}}{\laplace} G^\text{F}_2(x,x')\eqend{,}
\end{split}
\end{equations}
where $\laplace$ is the familiar Laplace operator in Euclidean space, $\Pi_{ij} \equiv \delta_{ij} - \frac{\del_i\del_j}{\laplace}$ is the transverse projector and $\frac{1}{\laplace}$ denotes the Green's function of the Laplace operator with boundary conditions that vanish at the spatial infinity. We recall that due to the gauge condition~(\ref{eq:gauge_conditions}), both the propagator for the inflaton perturbations and the correlator between $h_{\mu\nu}$ and $\phi^{(1)}$ are zero. The scalar propagators appearing in the right-hand side of Eqs.~(\ref{eq:graviton_propagator}) are defined by
\begin{equations}[eq:scalar_propagators_definition]
 \label{eq:G_H_eq}
 & [\del^2 - (n - 2)(Ha)(\eta)\del_\eta]G_\textrm{H}^\textrm{F}(x,x') = \frac{1}{a^{n - 2}(\eta)}\delta^{(n)}(x - x')\eqend{,}\\
 & [\del^2 - (n - 2 + 2\delta(\eta))(Ha)(\eta)\del_\eta]G_\textrm{Q}^\textrm{F}(x,x') = \frac{2}{(n - 2)(\epsilon a^{n - 2})(\eta)}\delta^{(n)}(x - x')\eqend{,}\\
 & [\del^2 - (n - 2)(Ha)(\eta)\del_\eta]G_2^\textrm{F}(x,x') = \laplace G_\textrm{H}^\textrm{F}(x,x')\eqend{,}\\
 \label{eq:D_H_eq}
 & \laplace D^\textrm{F}_\textrm{H}(x,x') = \del_\eta\del_{\eta'}G_\textrm{H}^\textrm{F}(x,x') - \frac{1}{a^{n - 2}(\eta)}\delta^{(n)}(x - x')\eqend{,}\\
 & \laplace D^\textrm{F}_\textrm{Q}(x,x') = \del_\eta\del_{\eta'}G_\textrm{Q}^\textrm{F}(x,x') - \frac{2}{(n - 2)(\epsilon a^{n - 2})(\eta)}\delta^{(n)}(x - x')\eqend{,}\\
 & \laplace D^\textrm{F}_2(x,x') = \del_\eta\del_{\eta'}G_2^\textrm{F}(x,x')\eqend{.}
\end{equations}
The graviton Wightman two-point function can be obtained from Eqs.~(\ref{eq:graviton_propagator}) and~(\ref{eq:scalar_propagators_definition}) simply by removing the terms containing the $\delta$-distribution in Eqs.~(\ref{eq:scalar_propagators_definition}). We will also need the spatial components of the propagator for the ghost field, which are given by
\begin{splitequation}
 \mathi G_{ij}^\textrm{F}(x,x') 
 &\equiv \theta(\eta - \eta')\expect{c_i(x)\bar{c}_j(x')}_0 - \theta(\eta' - \eta)\expect{\bar{c}_j(x')c_i(x)}_0\\
 & = \mathi \delta_{ij}G_\textrm{H}^\textrm{F}(x,x')\eqend{.}
\end{splitequation}

In the slow-roll approximation it is possible to simplify some of the expressions for the scalar propagators defined in Eq.~(\ref{eq:scalar_propagators_definition}), see Ref.~\cite{froeb_lima_cqg_2018}. They are
\begin{equations}[eq:G_2_D_2_propagators]
 G_2(x,x') & = -\frac{1}{2}\left[\eta\del_\eta + \eta'\del_{\eta'} - (n - 1) - (n - 2)\epsilon(\eta)\right]G_\mathrm{H}(x,x')\eqend{,}\\
 D_2(x,x') & = -\frac{1}{2}\left[\eta\del_\eta + \eta'\del_{\eta'} - (n - 3) - (n - 2)\epsilon(\eta)\right]D_\mathrm{H}(x,x')
\end{equations}
and
\begin{equation}\label{eq:G_H_D_H_relation}
 (\eta\del_{\eta'} + \eta'\del_\eta)G_\mathrm{H}(x,x') = \left\{\eta\del_\eta + \eta'\del_{\eta'} - 2(n - 2)[1 + \epsilon(\eta)]\right\}
 D_\mathrm{H}(x,x')\eqend{.}
\end{equation}
Due to the background and state spatial symmetries, it is convenient to express the scalar propagators in terms of their Fourier transform. The propagators then read as
\begin{equation}\label{eq:GFH_Fourier_transform}
 G^\textrm{F}_\textrm{H}(x,x') = \int\frac{d^{n - 1}p}{(2\pi)^{n - 1}}\tilde{G}^\textrm{F}_\textrm{H}(\eta,\eta',\vec{p})\mathe^{i\vec{p}\cdot(\vec{x} - \vec{x}')}\eqend{,}
\end{equation}
with
\begin{equation}\label{eq:GFH_Fourier_amplitude}
 \tilde{G}^\textrm{F}_\textrm{H}(\eta,\eta',\vec{p}) = \theta(\eta - \eta')\tilde{G}^+_\textrm{H}(\eta,\eta',\vec{p}) + \theta(\eta' - \eta)\tilde{G}^+_\textrm{H}(\eta',\eta,\vec{p})
\end{equation}
and the Wightman two-point function Fourier amplitude as
\begin{equation}\label{eq:Fourier_transform_G_H_slow_roll}
  \tilde{G}^+_\textrm{H}(\eta,\eta',\vec{p}) = -\mathi\frac{\pi}{4}\left\{[(1 - \epsilon)H](\eta)[(1 - \epsilon)H](\eta')\right\}^\frac{n - 2}{2}(\eta\eta')^\frac{n - 1}{2}\hankel{1}{\mu}(-p\eta)\hankel{2}{\mu}(-p\eta')\eqend{,}
\end{equation}
which corrects Eq.~(102) of Ref.~\cite{froeb_lima_cqg_2018}, while the Fourier transform of $D^\mathrm{F}_\mathrm{H}$ is 
\begin{equation}\label{eq:DFH_Fourier_transform}
 D^\textrm{F}_\textrm{H}(x,x') = \int\frac{d^{n - 1}p}{(2\pi)^{n - 1}}\tilde{D}^\textrm{F}_\textrm{H}(\eta,\eta',\vec{p})\mathe^{i\vec{p}\cdot(\vec{x} - \vec{x}')}\eqend{,}
\end{equation}
with
\begin{equation}\label{eq:DFH_Fourier_amplitude}
 \tilde{D}^\textrm{F}_\textrm{H}(\eta,\eta',\vec{p}) = \theta(\eta - \eta')\tilde{D}^+_\textrm{H}(\eta,\eta',\vec{p}) + \theta(\eta' - \eta)\tilde{D}^+_\textrm{H}(\eta',\eta,\vec{p})
\end{equation}
and the Wightman two-point function Fourier amplitude is
\begin{equation}\label{eq:Fourier_transform_D_H_slow_roll}
  \tilde{D}^+_\textrm{H}(\eta,\eta',\vec{p}) = \mathi\frac{\pi}{4}\left\{[(1 - \epsilon)H](\eta)[(1 - \epsilon)H](\eta')\right\}^\frac{n - 2}{2}(\eta\eta')^\frac{n - 1}{2}\hankel{1}{\mu - 1}(-p\eta)\hankel{2}{\mu - 1}(-p\eta')\eqend{,}
\end{equation}
which corrects Eq.~(104) of Ref.~\cite{froeb_lima_cqg_2018}. In Eqs.~(\ref{eq:Fourier_transform_G_H_slow_roll}) and~(\ref{eq:Fourier_transform_D_H_slow_roll}), we recall that $\hankel{1}{\alpha}(x)$ and $\hankel{2}{\alpha}(x)$ are the Hankel functions of first and second kinds, respectively, and order $\alpha$, $p \equiv \abs{\vec{p}}$, and the parameter $\mu$ depends on the slow-roll parameters as in Eq.~(\ref{eq:mu_definition}). The Fourier transform of the propagators $G^\mathrm{F}_\mathrm{Q}$ and $D^\mathrm{F}_\mathrm{Q}$ are
\begin{equation}\label{eq:GFQ_Fourier_transform}
 G^\textrm{F}_\textrm{Q}(x,x') = \int\frac{d^{n - 1}p}{(2\pi)^{n - 1}}\tilde{G}^\textrm{F}_\textrm{Q}(\eta,\eta',\vec{p})\mathe^{i\vec{p}\cdot(\vec{x} - \vec{x}')}\eqend{,}
\end{equation}
with
\begin{equation}\label{eq:GFQ_Fourier_amplitude}
 \tilde{G}^\textrm{F}_\textrm{Q}(\eta,\eta',\vec{p}) = \theta(\eta - \eta')\tilde{G}^+_\textrm{Q}(\eta,\eta',\vec{p}) + \theta(\eta' - \eta)\tilde{G}^+_\textrm{Q}(\eta',\eta,\vec{p})
\end{equation}
and the Wightman two-point function in Fourier space as
\begin{splitequation}\label{eq:Fourier_transform_G_Q}
  \tilde{G}^+_\textrm{Q}(\eta,\eta',\vec{p}) 
  & = -\mathi\frac{\pi}{2(n - 2)}\frac{\left\{[(1 - \epsilon)H](\eta)[(1 - \epsilon)H](\eta')\right\}^\frac{n - 2}{2}}{\sqrt{\epsilon(\eta)\epsilon(\eta')}}\\
  &\phantom{=} \times (\eta\eta')^\frac{n - 1}{2}\hankel{1}{\nu}(-p\eta)\hankel{2}{\nu}(-p\eta')\eqend{,}
\end{splitequation}
which corrects Eq.~(108) of Ref.~\cite{froeb_lima_cqg_2018}, and
\begin{equation}\label{eq:DFQ_Fourier_transform}
 D^\textrm{F}_\textrm{H}(x,x') = \int\frac{d^{n - 1}p}{(2\pi)^{n - 1}}\tilde{D}^\textrm{F}_\textrm{H}(\eta,\eta',\vec{p})\mathe^{i\vec{p}\cdot(\vec{x} - \vec{x}')}\eqend{,}
\end{equation}
with
\begin{equation}\label{eq:DFQ_Fourier_amplitude}
 \tilde{D}^\textrm{F}_\textrm{H}(\eta,\eta',\vec{p}) = \theta(\eta - \eta')\tilde{D}^+_\textrm{H}(\eta,\eta',\vec{p}) + \theta(\eta' - \eta)\tilde{D}^+_\textrm{H}(\eta',\eta,\vec{p})
\end{equation}
and
\begin{splitequation}\label{eq:Fourier_transform_D_Q}
  \tilde{D}^+_\textrm{Q}(\eta,\eta',\vec{p}) 
  & = \mathi\frac{\pi}{2(n - 2)}\frac{\left\{[(1 - \epsilon)H](\eta)[(1 - \epsilon)H](\eta')\right\}^\frac{n - 2}{2}}{\sqrt{\epsilon(\eta)\epsilon(\eta')}}\\
  &\phantom{=} \times (\eta\eta')^\frac{n - 1}{2}\hankel{1}{\nu - 1}(-p\eta)\hankel{2}{\nu - 1}(-p\eta')\eqend{,}
\end{splitequation}
which corrects Eq.~(110) of Ref.~\cite{froeb_lima_cqg_2018}. The parameter $\nu$ appearing above depends on the slow-roll parameters as in Eq.~(\ref{eq:nu_definition}).

%%%%%%%%%%%%%%%%%%%%%%%%%%%%%%%%%%%%%%%%%%%%%%%%%%%%%%%%%%%%%%%%%%%%%%%%%%%%%%%%%%%%%%%%%%%%%%%%%%%%%%%%%%%%%%%%%%%%%%%%%%%%%%%%%%%%%%%%%%%%%%%%%%%%%
\section{Analysis of the integral \boldmath\texorpdfstring{$J_{k, \alpha, \beta}$}{{\it J}}}                                                      %
\label{apdx:analysis_integral_J}                                                                                                                    %
%%%%%%%%%%%%%%%%%%%%%%%%%%%%%%%%%%%%%%%%%%%%%%%%%%%%%%%%%%%%%%%%%%%%%%%%%%%%%%%%%%%%%%%%%%%%%%%%%%%%%%%%%%%%%%%%%%%%%%%%%%%%%%%%%%%%%%%%%%%%%%%%%%%%%

In this appendix we analyse and compute the integral defined in Eq.~(\ref{eq:J_integral}). We start by performing the integration over the angular variables, which gives
\begin{equation}\label{eq:J_integral_angular_var_integrated}
 J_{k,\alpha,\beta} = \frac{1}{2^n\pi^\frac{n - 3}{2}\Gamma\left(\frac{n - 1}{2}\right)}\Re\int_0^\infty dq \textrm{H}_\alpha^{(1)}(q)\textrm{H}_\beta^{(2)}(q)q^{k + n - 2}\eqend{,}
\end{equation}
where $k \in \mathbb{Z}$ and $n, \alpha, \beta \in \mathbb{R}$.

As mentioned in Ref.~\cite{froeb_lima_cqg_2018}, the scalar propagators $G_\textrm{H}^\textrm{F}$ and $G_\textrm{Q}^\textrm{F}$ can be infrared (IR) divergent, depending on the values of the slow-roll parameters $\epsilon$ and $\delta$. Hence, it is worth analysing the IR behaviour of the integral $J_{k,\alpha,\beta}$ with respect to the value of its parameters. The limiting form of the Hankel function for small $q$~\cite{dlmf} gives
\begin{equation}
 J_{k,\alpha,\beta} = \dots + \textrm{cte}\times\int_0^\varepsilon dq q^{k + n - \alpha - \beta - 2}\eqend{,}
\end{equation}
which is IR finite if
\begin{equation}\label{eq:IR_finite_condition}
 k + n - \alpha - \beta - 1 > 0\eqend{.}
\end{equation}
It is easy to check that the condition~(\ref{eq:IR_finite_condition}) is satisfied by all terms in Eqs.~(\ref{eq:coincidence_limit_G_HQ_D_HQ}) and~(\ref{eq:coincidence_limit_G_HQ_D_HQ}) for all $\epsilon, |\delta| \ll 1$. For large $q$, however, we have that $\textrm{H}_\alpha^{(1)}(q)\textrm{H}_\beta^{(2)}(q)q^{k + n - 2} \sim q^{k + n - 3}$ and, thus, that $J_{k,\alpha,\beta}$ is divergent in the ultraviolet (UV) if $k + n - 3 \ge 0$, as expected. We will employ the dimensional regularisation to deal with the UV divergences.

In order to compute the integral in Eq.~(\ref{eq:J_integral_angular_var_integrated}) we use that
\begin{equation}
 \Re \textrm{H}_\alpha^{(1)}(q)\textrm{H}_\beta^{(2)}(q) = J_\alpha(q)J_\beta(q) + Y_\alpha(q)Y_\beta(q)\eqend{,} 
\end{equation}
where $J_\alpha(x)$ and $Y_\alpha(x)$ are the Bessel functions of first and second kind, respectively~\cite{dlmf}. It is convenient, however, to express the integrand in Eq.~(\ref{eq:J_integral_angular_var_integrated}) solely in terms of the Bessel function of first kind. To that end, we use that
\begin{equation}
 Y_\alpha(x) = \frac{\cos(\pi\alpha)J_\alpha(x) - J_{-\alpha}(x)}{\sin(\pi\alpha)}
\end{equation}
to obtain
\begin{splitequation}\label{eq:real_part_Hankel_functions}
 \Re \textrm{H}_\alpha^{(1)}(q)\textrm{H}_\beta^{(2)}(q) =
 & \frac{\sin(\pi\alpha)\sin(\pi\beta) + \cos(\pi\alpha)\cos(\pi\beta)}{\sin(\pi\alpha)\sin(\pi\beta)} J_ \alpha(q)J_\beta(q)\\ 
 & + \frac{1}{\sin(\pi\alpha)\sin(\pi\beta)}J_{-\alpha}(q)J_{-\beta}(q) - \frac{\cos(\pi\alpha)}{\sin(\pi\alpha)\sin(\pi\beta)}J_\alpha(q)J_{-\beta}(q)\\ 
 & - \frac{\cos(\pi\beta)}{\sin(\pi\alpha)\sin(\pi\beta)}J_{-\alpha}(q)J_\beta(q)\eqend{.}
\end{splitequation}

Next, we have from Eq.~(10.22.57) of Ref.~\cite{dlmf} that
\begin{equation}\label{eq:Bessel_J_integrated_1}
 \int_0^\infty dq J_\alpha(q)J_\beta{q}(q)q^{k + n - 2} = 
 \frac{2^{k + n - 2}\Gamma(2 - k - n)\Gamma\left(\frac{k + n + \alpha + \beta - 1}{2}\right)}{\Gamma\left(\frac{3 - k - n + \alpha + \beta}{2}\right) 
 \Gamma\left(\frac{3 - k - n + \alpha - \beta}{2}\right) \Gamma\left(\frac{3 - k - n - \alpha + \beta}{2}\right)}\eqend{,}
\end{equation}
provided that the conditions $k + n - 2 < 0$ and $k + n + \alpha + \beta - 1 > 0$ are satisfied. Note that the former condition is the convergence condition for the UV, while the latter can be obtained from the condition~(\ref{eq:IR_finite_condition}) for convergence in the IR. We then use the reflection formula for the $\Gamma$-functions in Eq.~(\ref{eq:Bessel_J_integrated_1}) to cast the right-hand side of that expression in the form
\begin{splitequation}\label{eq:Bessel_J_integrated_2}
 & \int_0^\infty dq J_\alpha(q)J_\beta{q}(q)q^{k + n - 2} = \\
 & \frac{2^{k + n - 2}}{\pi^2}\frac{\cos\left[\frac{\pi}{2}(k + n + \alpha - \beta)\right] \cos\left[\frac{\pi}{2}(k + n - \alpha + \beta)\right] 
 \cos\left[\frac{\pi}{2}(k + n - \alpha - \beta)\right]}{\sin[\pi(k + n)]\Gamma(k + n - 1)}\\
 &\times \Gamma\left(\frac{k + n + \alpha + \beta - 1}{2}\right) \Gamma\left(\frac{k + n + \alpha - \beta - 1}{2}\right)
 \Gamma\left(\frac{k + n - \alpha + \beta - 1}{2}\right)\\
 &\times \Gamma\left(\frac{k + n - \alpha - \beta - 1}{2}\right)\eqend{.}
\end{splitequation}
Finally, with the aid of Eq.~(\ref{eq:Bessel_J_integrated_2}) and the change
\begin{splitequation}
 \alpha & = \mu - a\eqend{,}\\
 \beta  & = \mu - b\eqend{,}
\end{splitequation}
with $a, b \in \mathbb{Z}$, we can express the integral in Eq.~(\ref{eq:J_integral_angular_var_integrated}) as
\begin{splitequation}
 J_{k,\mu - a, \mu - b} =
 & (-1)^{a + b + k}\frac{2^{k - 1}}{\pi^\frac{n + 1}{2}}\frac{\cos(\pi\mu)\cos\left[\frac{\pi}{2}(k + n + a + b)\right] \Gamma\left(\frac{k + n + a - b - 1}{2}\right)\Gamma\left(\frac{k + n - a + b - 1}{2}\right)}{\Gamma(k + n - 1)\Gamma\left(\frac{n - 1}{2}\right)\sin[\pi(n - 4)]}\\
 &\times \Gamma\left(\frac{k + n + a + b - 1}{2} - \mu\right)\Gamma\left(\frac{k + n - a - b - 1}{2} + \mu\right)\eqend{,}
\end{splitequation}
after performing some manipulations involving trigonometric identities. The expression above diverges as $n \to 4$, as expected. We provide a list of the values of $J_{k,\mu - a, \mu - b}$ needed for this paper for a given $\mu$. They are
\begin{equations}[eq:J_integral_values]
& J_{0,\mu - 1, \mu - 1} = -A_\mu^{(n)}n \eqend{,}\\
& J_{1,\mu, \mu - 1} = -A_\mu^{(n)}n\left(\frac{n - 3}{2} + \mu\right) \eqend{,}\\
& J_{2,\mu - 1, \mu - 1} = A_\mu^{(n)}(n - 1)\left(\frac{n + 1}{2} - \mu\right)\left(\frac{n - 3}{2} + \mu\right) \eqend{,}\\
& J_{2,\mu, \mu} = -A_\mu^{(n)}(n - 1)\left(\frac{n - 1}{2} + \mu\right)\left(\frac{n - 3}{2} + \mu\right) \eqend{,}\\
& J_{3,\mu, \mu - 1} = A_\mu^{(n)}(n - 1)\left(\frac{n + 1}{2} - \mu\right)\left(\frac{n - 1}{2} + \mu\right)\left(\frac{n - 3}{2} + \mu\right) \eqend{,}\\
& J_{4,\mu - 1, \mu - 1} = -A_\mu^{(n)}\frac{n^2 - 1}{n + 2}\left(\frac{n + 3}{2} - \mu\right)\left(\frac{n + 1}{2} - \mu\right)\left(\frac{n + 1}{2} + \mu\right)\\
&\qquad\qquad\qquad\times\left(\frac{n - 1}{2} + \mu\right)\left(\frac{n - 3}{2} + \mu\right) \eqend{,}\\
& J_{4,\mu, \mu} = A_\mu^{(n)}\frac{n^2 - 1}{n + 2}\left(\frac{n + 1}{2} - \mu\right)\left(\frac{n + 1}{2} + \mu\right)\left(\frac{n - 1}{2} + \mu\right)\left(\frac{n - 3}{2} + \mu\right) \eqend{,}
\end{equations}
where we have defined
\begin{equation}\label{eq:A_mu}
 A_\mu^{(n)} \equiv \frac{\cos\left(\frac{\pi}{2}n\right)\cos(\pi\mu)\Gamma\left(\frac{n + 1}{2} - \mu\right)\Gamma\left(\frac{n - 3}{2} + \mu\right)}{2^n\pi^\frac{n}{2}\Gamma\left(\frac{n + 2}{2}\right)\sin[\pi(n - 4)]}\eqend{.}
\end{equation}

In terms of that integral, the coincidence limit of the derivatives of the scalar propagators we need in this paper are 
\begin{equations}[eq:coincidence_limit_G_HQ_D_HQ]
 \mathi \lim_{x' \to x}(\del_\eta + \del_{\eta'})G^\textrm{F}_\textrm{H}(x,x') &= - 2[(1 - \epsilon)H]^{n - 1}aJ_{1,\mu,\mu - 1}\eqend{,}\\
 \mathi \lim_{x' \to x}\laplace G^\textrm{F}_\textrm{H}(x,x') &= - [(1 - \epsilon)H]^na^2J_{2,\mu,\mu}\eqend{,}\\
 \mathi \lim_{x' \to x}\del_\eta\del_{\eta'}G^\textrm{F}_\textrm{H}(x,x') &= [(1 - \epsilon)H]^na^2J_{2,\mu - 1,\mu - 1}\eqend{,}\\
 \mathi \lim_{x' \to x}(\del_\eta + \del_{\eta'})\laplace G^\textrm{F}_\textrm{H}(x,x') &= 2[(1 - \epsilon)H]^{n + 1}a^3J_{3,\mu,\mu - 1}\eqend{,}\\
 \mathi \lim_{x' \to x}D^\textrm{F}_\textrm{H}(x,x') &= - [(1 - \epsilon)H]^{n - 2}J_{0,\mu - 1,\mu - 1}\eqend{,}\\
 \mathi \lim_{x' \to x}(\del_\eta + \del_{\eta'})\GFQ &= - \frac{4}{n - 2}\frac{[(1 - \epsilon)H]^{n - 1}}{\epsilon}aJ_{1,\nu,\nu - 1}\eqend{,}\\
 \mathi \lim_{x' \to x}\laplace \GFQ &= - \frac{2}{n - 2}\frac{[(1 - \epsilon)H]^n}{\epsilon}a^2J_{2,\nu,\nu}\eqend{,}\\
 \mathi \lim_{x' \to x}\del_\eta\del_{\eta'}\GFQ &= \frac{2}{n - 2}\frac{[(1 - \epsilon)H]^n}{\epsilon}a^2J_{2,\nu - 1,\nu - 1}\eqend{,}\\
 \mathi \lim_{x' \to x}(\del_\eta + \del_{\eta'})\laplace \GFQ &= \frac{4}{n - 2}\frac{[(1 - \epsilon)H]^{n + 1}}{\epsilon}a^3J_{3,\mu,\mu - 1} \eqend{,}\\
 \mathi \lim_{x' \to x}\DFQ &= - \frac{2}{n - 2}\frac{[(1 - \epsilon)H]^{n - 2}}{\epsilon}J_{0,\nu - 1,\nu - 1}\eqend{,}\\
 \mathi \lim_{x' \to x}\del_\eta\del_{\eta'}\laplace\GFQ & = - \frac{2}{n - 2}\frac{[(1 - \epsilon)H]^{n + 2}}{\epsilon}a^4 J_{4,\nu - 1,\nu - 1}\eqend{,}\\
 \mathi \lim_{x' \to x}\laplace^2\GFQ & = \frac{2}{n - 2}\frac{[(1 - \epsilon)H]^{n + 2}}{\epsilon}a^4 J_{4,\nu,\nu}\eqend{.}
\end{equations}

%%%%%%%%%%%%%%%%%%%%%%%%%%%%%%%%%%%%%%%%%%%%%%%%%%%%%%%%%%%%%%%%%%%%%%%%%%%%%%%%%%%%%%%%%%%%%%%%%%%%%%%%%%%%%%%%%%%%%%%%%%%%%%%%%%%%%%%%%%%%%%%%%%%%%
\section{Coincidence limits in the three-gravitons interaction terms}                                                                               %
\label{apdx:coincidence_limit_three_gravitons}                                                                                                      %
%%%%%%%%%%%%%%%%%%%%%%%%%%%%%%%%%%%%%%%%%%%%%%%%%%%%%%%%%%%%%%%%%%%%%%%%%%%%%%%%%%%%%%%%%%%%%%%%%%%%%%%%%%%%%%%%%%%%%%%%%%%%%%%%%%%%%%%%%%%%%%%%%%%%%

In this appendix we provide the intermediate steps of the computation of the three-gravitons interaction terms contribution to the loop correction of the local expansion rate.

\subsection{The \texorpdfstring{$V$}{V}-tensor term}
%%%%%%%%%%%%%%%%%%%%%%%%%%%%%%%%%%%%%%%%%%%%%%%%%%%%%%%%%%%%%%%%%%%%%%%%%%%%%%%%%%%%%%%%%%%%%%%%%%%%%%%%%%%%%%%%%%%%%%%%%%%%%%%%%%%%%%%%%%%%%%%%%%%%%
We start by substituting Eq.~(\ref{eq:F_ij_slow_roll}) in Eq.~(\ref{eq:V_contribution_simplified}) and integrating by parts. The result is 
\begin{splitequation}\label{eq:V_contribution_simplified_slow_roll}
 \mathi \expect{H^{(1)}(x)S^{(1)}_{\textrm{G}, V}}_0
 & = \frac{\mathi(n - 2)\epsilon(\eta)}{8(n - 1)a(\eta)}\int_{-\infty}^\eta d\eta'\int d^{n - 1}x'[\del_\eta G_\mathrm{Q}^+(x,x') - \del_\eta G_\mathrm{Q}^+(x',x)]\\
 &\phantom{=}\qquad \times \left\{(Ha^{n - 1})(\eta')\delta_{ij}V^{ij\mu\nu\rho\sigma}\lim_{y,y' \to x'}\del_\rho G^\mathrm{F}_{0\sigma\mu\nu}(y,y') \right.\\
 & \phantom{=}\qquad\qquad \left. - \delta_{ij}V^{\alpha\beta ij 0\sigma}\del_{\eta'}\left[(Ha^{n - 1})(\eta')\lim_{y,y' \to x'} G^\mathrm{F}_{\alpha\beta0\sigma}(y,y')\right]\right\}\eqend{,}
\end{splitequation}
where
\begin{equation}\label{eq:V_tensor}
 V^{\alpha\beta\mu\nu\rho\sigma} \equiv \eta^{\alpha\beta}\eta^{\mu\nu}\eta^{\rho\sigma} - 2\eta^{\alpha\mu}\eta^{\beta\nu}\eta^{\rho\sigma}
 - 2\eta^{\alpha\rho}\eta^{\mu\nu}\eta^{\beta\sigma}\eqend{.}
\end{equation}
Next, we compute the coincidence limits appearing in the expression above. They are 
\begin{splitequation}
 & \delta_{ij}V^{ij\mu\nu\rho\sigma}\lim_{x' \to x} \del_\rho G^\mathrm{F}_{0\sigma\mu\nu}(x,x')\\
 & = \lim_{x' \to x}\left[(n - 1)\del_\eta G^\mathrm{F}_{0000}(x,x') - (n - 3)\del^iG^\mathrm{F}_{0i00}(x,x') - (n - 3)\del_\eta G^{\mathrm{F}\phantom{0}k}_{00\phantom{k}k}(x,x')\right.\\ 
 & \phantom{=}\left. + (n - 5)\del^iG^{\mathrm{F}\phantom{i}\,k}_{0i\phantom{k}k}(x,x')\right]\\ 
 & = \lim_{x' \to x}\left\{-\frac{(n - 1)(n - 5)}{2(n - 2)}(\del_\eta + \del_{\eta'})\GFH - \frac{n - 5}{(1 - \epsilon)Ha}(\del_\eta\del_{\eta'} + \laplace)\GFH \right.\\
 & \phantom{=} -\frac{n - 1}{4}[2(n - 1)(n - 3) - (3n - 11)\epsilon + 4(n - 3)\delta](\del_\eta + \del_{\eta'})\GFQ\\
 & \phantom{=} -\frac{1}{2Ha}[2(n - 1)^2 - (3n - 5)\epsilon + 4(n - 1)\delta]\del_\eta\del_{\eta'}\GFQ + \frac{(n - 1)^2}{2Ha}\laplace \GFQ\\
 & \phantom{=} \left. + \frac{n - 1}{4(Ha)^2}(\del_\eta + \del_{\eta'})\laplace\GFQ\right\}\\
 & = \frac{\mathi[(1 - \epsilon)H]^{n - 1}a}{(n - 2)\epsilon}\left\{-(n - 5)(n - 1)\epsilon J_{1,\mu,\mu - 1} + (n - 5)(n - 2)\epsilon(J_{2,\mu - 1,\mu - 1} - J_{2,\mu,\mu})\phantom{^2}\right.\\
 & \phantom{=} - (n - 1)[2(n - 1)(n - 3) - (3n - 11)\epsilon + 4(n - 3)\delta]J_{1,\nu,\nu - 1}\\ 
 & \phantom{=} + [2(n - 1)^2 - (3n - 5)\epsilon + 4(n - 1)\delta](1 - \epsilon)J_{2,\nu - 1,\nu - 1} + (n - 1)^2(1 - \epsilon)J_{2,\nu,\nu}\\
 & \phantom{=} \left. - (n - 1)(1 - \epsilon)^2J_{3,\nu,\nu - 1}\right\}\eqend{,}
\end{splitequation}
and
\begin{splitequation}
 & \delta_{ij} V^{\alpha\beta ij0\sigma}\lim_{x' \to x}G^\mathrm{F}_{\alpha\beta0\sigma}(x,x')\\
 & = \lim_{x' \to x}\left[2(n - 1) G^{\mathrm{F}k}_{\phantom{\mathrm{F}k}0k0}(x,x') - (n - 3)G^{\mathrm{F}k}_{\phantom{\mathrm{F}k}k00}(x,x')
  - (n - 1)G^\mathrm{F}_{000}(x,x')\right]\\
 & = \lim_{x' \to x}\left\{\frac{(n - 1)^2}{(1 - \epsilon)Ha}(\del_\eta + \del_{\eta'})\GFH - \frac{(n - 1)[(n - 3)(n^2 - 3n + 3) + (n - 2)^2]}{n - 2}\right.\\
 & \phantom{=} \times \DFH + \frac{n - 1}{2Ha}(n - 3 + \epsilon)(\del_\eta + \del_{\eta'})\GFQ - \frac{n - 1}{2(Ha)^2}(2\del_\eta\del_{\eta'} + \laplace)\GFQ\\ 
 & \phantom{=} \left. - \frac{(n - 1)\epsilon^2}{2}\DFQ\right\}\\
 & = \frac{\mathi(n - 1)[(1 - \epsilon)H]^{n - 2}}{(n - 2)\epsilon}\left\{2(n - 2)(n - 1)\epsilon J_{1,\mu,\mu - 1} - [(n - 3)(n^2 - 3n + 3)
 \right.\\
 & \phantom{=} + (n - 2)^2\epsilon]\epsilon J_{0,\mu - 1,\mu - 1} + 2(n - 3 + \epsilon)(1 - \epsilon)J_{1,\nu,\nu - 1} + (1 - \epsilon)^2(2J_{2,\nu - 1,\nu - 1} - J_{2,\nu,\nu})\\
 & \phantom{=} \left. - \epsilon^2 J_{0,\nu - 1,\nu - 1}\right\}\eqend{.}
\end{splitequation}
We then substituting these limits into Eq.~(\ref{eq:V_contribution_simplified_slow_roll}) and pull the terms that vary in time at orders higher than one in the slow-roll parameters out of the integral. This results in the expression of Eq.~(\ref{eq:V_contribution_J}).

\subsection{The \texorpdfstring{$U$}{U}-tensor term}
%%%%%%%%%%%%%%%%%%%%%%%%%%%%%%%%%%%%%%%%%%%%%%%%%%%%%%%%%%%%%%%%%%%%%%%%%%%%%%%%%%%%%%%%%%%%%%%%%%%%%%%%%%%%%%%%%%%%%%%%%%%%%%%%%%%%%%%%%%%%%%%%%%%%%
We use the expressions for the components of $F_{\mu\nu}$, given in Eqs.~(\ref{eq:F_00_slow_roll}), (\ref{eq:F_ij_slow_roll}) and~(\ref{eq:F_0i_slow_roll}), in Eq.~(\ref{eq:U_contribution}), discard the terms in the resulting expressions involving total spatial derivatives, and integrate by parts. This leads to
\begin{splitequation}\label{eq:U_simplified_slow_roll}
 \mathi \expect{H^{(1)}(x)S^{(1)}_{\textrm{G}, U}}_0
 & = \frac{\mathi}{16}\frac{\epsilon(\eta)}{(n - 1)a(\eta)}\delta_{ij}\int_{-\infty}^\eta d\eta' \int d^{n - 1}x'a^{n - 2}(\eta')\\
 & \phantom{=} \times \left[\del_\eta G^+_\mathrm{Q}(x,x') - \del_\eta G^+_\mathrm{Q}(x',x)\right]\left\{U^{\alpha\beta ij\mu\nu\rho\sigma}\lim_{y,y' \to x'} \del_\alpha\del_\beta'G^\mathrm{F}_{\mu\nu\rho\sigma}(y,y')\phantom{\frac{1^2}{1}}\right.\\
 & \phantom{=} \left. - \frac{1}{a^{n - 2}(\eta')}\frac{d}{d\eta'}\left[a^{n - 2}(\eta')(U^{0\beta\rho\sigma ij\mu\nu} + U^{\beta0 \rho\sigma \mu\nu ij})\lim_{y,y' \to x'} \del_\beta G^\mathrm{F}_{\rho\sigma\gamma\delta}(y,y')\right]\right\}\eqend{,}
\end{splitequation}
with
\begin{splitequation}\label{eq:U_tensor}
 U^{\alpha\beta\gamma\delta\mu\nu\rho\sigma} \equiv
 &\ 2\eta^{\mu\rho}\eta^{\alpha\sigma}\eta^{\nu\beta}\eta^{\gamma\delta} - 4\eta^{\alpha\sigma}\eta^{\nu\beta}\eta^{\gamma\mu}\eta^{\delta\rho}
    - 4\eta^{\mu\rho}\eta^{\nu\beta}\eta^{\alpha\gamma}\eta^{\sigma\delta} - 4\eta^{\mu\rho}\eta^{\alpha\sigma}\eta^{\gamma\nu}\eta^{\delta\beta}\\
 & - 2\eta^{\mu\nu}\eta^{\alpha\sigma}\eta^{\rho\beta}\eta^{\gamma\delta} + 4\eta^{\alpha\sigma}\eta^{\rho\beta}\eta^{\gamma\mu}\eta^{\delta\nu}
   + 4\eta^{\mu\nu}\eta^{\rho\beta}\eta^{\alpha\gamma}\eta^{\sigma\delta} + 4\eta^{\mu\nu}\eta^{\alpha\sigma}\eta^{\gamma\rho}\eta^{\delta\beta}\\
 & - \eta^{\mu\rho}\eta^{\alpha\beta}\eta^{\nu\sigma}\eta^{\gamma\delta} + 2\eta^{\alpha\beta}\eta^{\nu\sigma}\eta^{\gamma\mu}\eta^{\delta\rho}
   + 2\eta^{\mu\rho}\eta^{\nu\sigma}\eta^{\alpha\gamma}\eta^{\beta\delta} + 2\eta^{\mu\rho}\eta^{\alpha\beta}\eta^{\gamma\nu}\eta^{\delta\sigma}\\
 & + \eta^{\alpha\beta}\eta^{\mu\nu}\eta^{\gamma\delta}\eta^{\rho\sigma} - 2\eta^{\alpha\beta}\eta^{\rho\sigma}\eta^{\gamma\mu}\eta^{\delta\nu}
   - 2\eta^{\mu\nu}\eta^{\rho\sigma}\eta^{\alpha\gamma}\eta^{\beta\delta} - 2\eta^{\alpha\beta}\eta^{\mu\nu}\eta^{\gamma\rho}\eta^{\delta\sigma}
\end{splitequation}
We then turn to the computation of the coincidence limits of the Feynman graviton propagator. Let us start by calculating
\begin{splitequation}\label{eq:U_del_del_G}
 & \delta_{ij}U^{\alpha\beta ij \mu\nu\rho\sigma}\lim_{x' \to x} \del_\alpha\del_\beta'G^\mathrm{F}_{\mu\nu\rho\sigma}(x,x') \\
 & = \lim_{x' \to x}\left\{(n - 5)\del_\eta\del_{\eta'}[G^{\mathrm{F}ij}_{\phantom{\mathrm{F}ij}ij}(x,x') - G^{\mathrm{F}i\phantom{i}j}_{\phantom{\mathrm{F}i}i\phantom{j}j}(x,x')] + 2(n - 3)\del_\eta\del^i[G^\mathrm{F}_{000i}(x,x')\right.\\
 & \phantom{=} - G^\mathrm{F}_{0i00}(x,x')] + 2(n - 5)\del_\eta\del^i[2G^{\mathrm{F}j}_{\phantom{\mathrm{F}j}ij0}(x,x') - G^{\mathrm{F}j}_{\phantom{\mathrm{F}j}ji0}(x,x') - G^{\mathrm{F}\phantom{0i}j}_{\phantom{\mathrm{F}}0i\phantom{j}j}(x,x')]\\ 
 & \phantom{=} + 2(n - 5)\del^i\del^j[G^\mathrm{F}_{0i0j}(x,x') - G^\mathrm{F}_{00ij}(x,x')] + 2(n - 7)\del^i\del^j[G^{\mathrm{F}k}_{\phantom{\mathrm{F}k}kij}(x,x') - G^{\mathrm{F}k}_{\phantom{\mathrm{F}k}ikj}(x,x')]\\ 
 & \phantom{=} \left.+ 2(n - 5)\laplace [G^{\mathrm{F}\phantom{00}i}_{\phantom{\mathrm{F}}00\phantom{i}i}(x,x') - G^{\mathrm{F}i}_{\phantom{\mathrm{F}i}0i0}(x,x')] + (n - 7)\laplace[G^{\mathrm{F}ij}_{\phantom{\mathrm{F}ij}ij}(x,x') - G^{\mathrm{F}i\phantom{i}j}_{\phantom{\mathrm{F}i}i\phantom{j}j}(x,x')]\right\}
\end{splitequation}
Using Eq.~(\ref{eq:graviton_propagator}), we can express this contraction in terms of the scalar propagators as
\begin{splitequation}
 & \delta_{ij} U^{\alpha\beta ij \mu\nu \rho\sigma}\lim_{x' \to x}\del_\alpha\del'_\beta G^\mathrm{F}_{\mu\nu\rho\sigma}(x,x')\\
 & = \lim_{x' \to x}\left\{ - (n - 5)(n - 1)Ha(\del_\eta + \del_{\eta'})\GFH + \frac{n - 5}{(n - 2)(1 - \epsilon)}[(n - 1)(14 - 8n + n^2)\right.\\ 
 & \phantom{=} - (2 + 6n - 5n^2 + n^3)\epsilon + 2(n - 2)^2\epsilon^2]\del_\eta\del_{\eta'}\GFH - \frac{1}{n - 2}(10 + 40n - 41n^2\\
 & \phantom{=} + 12n^3 - n^4)\laplace\GFH + \frac{4(n - 5)}{(1 - \epsilon)Ha}(\del_\eta + \del_{\eta'})\laplace \GFH\\ 
 & \phantom{=} + \frac{(n - 5)(n - 2)(n - 1)\epsilon Ha}{2}(\del_\eta + \del_{\eta'})\GFQ - [(n - 5)(n - 2)(n - 1)\\ 
 & \phantom{=} - (6 + n - n^2)\epsilon + 2(n - 3)\epsilon\delta - (n - 3)\epsilon^2]\del_\eta\del_{\eta'}\GFQ + (37 - 35n + 11n^2\\ 
 & \phantom{=} - n^3)\laplace \GFQ - \frac{(n - 3)(n - 10) - 2(n - 3)\delta}{2Ha}(\del_\eta + \del_{\eta'})\laplace \GFQ \\
 & \phantom{=} \left. - \frac{n - 3}{(Ha)^2}(\del_\eta\del_{\eta'} + \laplace)\laplace\GFQ\right\}\eqend{.}
 \end{splitequation}
 We then substitute the coincidence limits of the derivatives of the scalar propagators found in Eq.~(\ref{eq:coincidence_limit_G_HQ_D_HQ}) in the expression above. This yields
\begin{splitequation}\label{eq:U_del_del_G_slow_roll}
 & \delta_{ij} U^{\alpha\beta ij \mu\nu \rho\sigma}\lim_{x' \to x}\del_\alpha\del'_\beta G^\mathrm{F}_{\mu\nu\rho\sigma}(x,x')\\
 & = \frac{\mathi[(1 - \epsilon)H]^na^2}{(n - 2)\epsilon}\left\{-\frac{2(n - 5)(n - 1)\epsilon}{1 - \epsilon}J_{1,\mu,\mu - 1} - \frac{(n - 5)(n - 1)(14 - 8n + n^2)\epsilon}{1 - \epsilon}\right.\\
 & \phantom{=} \times J_{2,\mu - 1,\mu - 1}- (10 + 40n - 41n^2 + 12n^3 - n^4)\epsilon J_{2,\mu,\mu} - 8(n - 5)(n - 2)\epsilon J_{3,\mu,\mu - 1}\\
 & \phantom{=} + \frac{2(n - 5)(n - 2)(n - 1)\epsilon}{1 - \epsilon}J_{1,\nu,\nu - 1} + [2(n - 5)(n - 2)(n - 1) - 2(6 + n - n^2)\epsilon]\\
 & \phantom{=} \times J_{2,\nu - 1,\nu - 1} + 2(37 - 35n + 11n^2 - n^3)J_{2,\nu,\nu} + 2[(n - 10)(n - 3)(1 - \epsilon)\\ 
 & \left.\phantom{\frac{1^2}{1}} - 2(n - 3)\delta]J_{3,\nu,\nu - 1} + 2(n - 3)(1 - \epsilon)^2(J_{4,\nu,\nu} - J_{4,\nu - 1,\nu - 1})\right\}\eqend{.}
\end{splitequation}
The other coincidence limit is given by
\begin{splitequation}\label{eq:U_del_G}
 & \delta_{ij}(U^{0\beta\rho\sigma ij\mu\nu} + U^{\beta0 \rho\sigma \mu\nu ij})\lim_{x' \to x} \del_\beta G^\mathrm{F}_{\mu\nu\rho\sigma}(x,x')\\
 & = \lim_{x' \to x}\left\{\del_\eta[4(n - 3)G^{\mathrm{F}ij}_{\phantom{\mathrm{F}ij}ij}(x,x') - 2(n - 4)G^{\mathrm{F}i\phantom{i}j}_{\phantom{\mathrm{F}i}i\phantom{j}j}(x,x') - 2(n - 2)G^{\mathrm{F}i}_{\phantom{\mathrm{F}i}i00}(x,x')]\right.\\
 & \phantom{=} \left. + \del^i[2(n - 1)G^\mathrm{F}_{0i00}(x,x') - 2(n - 1)G^{\mathrm{F}\phantom{0i}j}_{\phantom{\mathrm{F}}0i\phantom{j}j}(x,x') - 8G^{\mathrm{F}j}_{\phantom{\mathrm{F}j}ij0}(x,x')]\right\}\eqend{,}
\end{splitequation}
which we again express in terms of the scalar propagators, resulting in
\begin{splitequation}
 & \delta_{ij}(U^{0\beta\rho\sigma ij \mu\nu} + U^{\beta0\rho\sigma\mu\nu ij})\lim_{x' \to x} \del_\beta G^\mathrm{F}_{\mu\nu\rho\sigma}(x,x')\\
 & = \lim_{x' \to x}\left[\frac{-7 - 8n + 21n^2 -12n^3 + 2n^4}{n - 2}(\del_\eta + \del_{\eta'})\GFH\right.\\
 & \phantom{=} - \frac{18 - 14n + 4n^2}{(1 - \epsilon)Ha}(\del_\eta\del_{\eta'} + \laplace)\GFH + \left(10 - 17n + 8n^2 - n^3 \phantom{\frac{1^2}{1}}\right.\\
 & \phantom{=} \left. + \frac{3 + 2n - n^2}{2}\epsilon\right)(\del_\eta + \del_{\eta'})\GFQ + \frac{2(n - 2)(n - 1) - (n - 1)\epsilon}{Ha}\del_\eta\del_{\eta'}\GFQ\\ 
 & \phantom{=} \left. - \frac{3 + 2n - n^2}{Ha}\laplace\GFQ + \frac{n - 1}{(Ha)^2}(\del_\eta + \del_{\eta'})\laplace\GFQ\right]\eqend{.}
\end{splitequation}
Their coincidence limit then yields
\begin{splitequation}\label{eq:U_del_G_slow_roll}
 & \delta_{ij}(U^{0\beta\rho\sigma ij \mu\nu} + U^{\beta0\rho\sigma\mu\nu ij})\lim_{x' \to x} \del_\beta G^\mathrm{F}_{\mu\nu\rho\sigma}(x,x')\\
 & = \frac{2\mathi[(1 - \epsilon)H]^{n - 1}a}{(n - 2)\epsilon}\left\{(-7 - 8n + 21n^2 - 12n^3 + 2n^4)\epsilon J_{1,\mu,\mu - 1}\right.\\
 & \phantom{=} + (n - 2)(9 - 7n + 2n^2)\epsilon(J_{2,\mu - 1,\mu - 1} - J_{2,\mu,\mu}) - [2(n - 5)(n - 2)(n - 1)\\ 
 & \phantom{=} + (n + 1)(n - 3)\epsilon]J_{1,\nu,\nu - 1} - (n - 1)(2n - 4 - \epsilon)(1 - \epsilon)J_{2,\nu - 1,\nu - 1}\\ 
 & \phantom{=} \left. + (n - 3)(n + 1)(1 - \epsilon)J_{2,\nu,\nu} - (n - 1)(1 - \epsilon)^2J_{3,\nu,\nu - 1}\right\}\eqend{.}
\end{splitequation}
Finally, we substitute Eqs.~(\ref{eq:U_del_del_G_slow_roll}) and~(\ref{eq:U_del_G_slow_roll}) into Eq.~(\ref{eq:U_simplified_slow_roll}), resulting in Eq.~(\ref{eq:U_contribution_J}).

%%%%%%%%%%%%%%%%%%%%%%%%%%%%%%%%%%%%%%%%%%%%%%%%%%%%%%%%%%%%%%%%%%%%%%%%%%%%%%%%%%%%%%%%%%%%%%%%%%%%%%%%%%%%%%%%%%%%%%%%%%%%%%%%%%%%%%%%%%%%%%%%%%%%%
\section{Backreaction in the constant-\boldmath\texorpdfstring{$\epsilon$}{\textepsilon} case}                                                    %
\label{apdx:cte_epsilon}                                                                                                                            %
%%%%%%%%%%%%%%%%%%%%%%%%%%%%%%%%%%%%%%%%%%%%%%%%%%%%%%%%%%%%%%%%%%%%%%%%%%%%%%%%%%%%%%%%%%%%%%%%%%%%%%%%%%%%%%%%%%%%%%%%%%%%%%%%%%%%%%%%%%%%%%%%%%%%%

In this appendix we show the corrected expressions for the one-loop contributions computed in Ref.~\cite{froeb_cqg_2019}, as well as the correct expressions for the scalar propagators of Ref.~\cite{froeb_lima_cqg_2018} needed in that calculation. We start by the latter. In the constant-$\epsilon$ case all we need are the scalar propagators $G_\mathrm{H}^\mathrm{F}$ and $D_\mathrm{H}^\mathrm{F}$, defined in Eqs.~(\ref{eq:scalar_propagators_definition}). The Fourier amplitude of the corresponding Wightman two-point functions are
\begin{equations}
\tilde{G}^+_\textrm{H}(\eta,\eta',\vec{p}) & = -\mathi\frac{\pi}{4}(1 - \epsilon)^{n - 2}\left[H(\eta)H(\eta')\right]^\frac{n - 2}{2}(\eta\eta')^\frac{n - 1}{2}\hankel{1}{\mu}(-p\eta)\hankel{2}{\mu}(-p\eta')\eqend{,}\\
\tilde{D}^+_\textrm{H}(\eta,\eta',\vec{p}) 
  &= \mathi\frac{\pi}{4}(1 - \epsilon)^{n - 2}\left[H(\eta)H(\eta')\right]^\frac{n - 2}{2}(\eta\eta')^\frac{n - 1}{2}\hankel{1}{\mu - 1}(-p\eta)\hankel{2}{\mu - 1}(-p\eta')\eqend{,}
\end{equations}
respectively. The contribution coming from pure second-order term, counter-terms, ghost, and three-graviton interaction terms are
\begin{equation}
 \expect{H^{(2)}(x)}_0 = -H^{n - 1}C_2(n,\epsilon)\eqend{,}
\end{equation}
with
\begin{splitequation}\label{eq:C_2}
 C_2(n,\epsilon) 
 & = \frac{A_\mu^{(n)}}{16(n - 2)}\frac{(1 - \epsilon)^{n - 2}}{\epsilon}\left[4n(n^2 + n - 6) + 2(8 + 28n - 9n^2 - 7n^3 + 2n^4)\epsilon\right.\\
 & \phantom{=} \left. + 8(2n^2 - 4n - 1)\epsilon^2 - n(n^2 - 4)\epsilon^3\right]\eqend{,}
\end{splitequation}
\begin{equation}\label{eq:CT_cte_epsilon}
 \mathi \expect{H^{(1)}(x)S^{(1)}_\mathrm{G,CT}}_0 = \frac{H}{2}\delta_V\eqend{,} 
\end{equation}
\begin{equation}
 \mathi \expect{H^{(1)}(x)S_{\textrm{GH},\textrm{eff}}^{(1)}}_0 = -H^{n - 1}C_\mathrm{GH}(n,\epsilon)\eqend{,}
\end{equation}
with
\begin{equation}\label{eq:C_GH}
 C_\mathrm{GH}(n,\epsilon) = \frac{nA^{(n)}_\mu}{2}(1 - \epsilon)^{n - 2}(2 - \epsilon)\eqend{,}
\end{equation}
\begin{equation}
 \mathi \expect{H^{(1)}(x)S^{(1)}_{\textrm{G}, V}}_0 = -H^{n - 1}C_{\mathrm{G},V}(n,\epsilon)\eqend{,}
\end{equation}
with
\begin{splitequation}\label{eq:C_G_V}
 C_{\mathrm{G},V}(n,\epsilon) 
 & = \frac{(1 - \epsilon)^{n - 3}A^{(n)}_\mu}{32(n - 1)^2(n - 2)\epsilon}\left[8(2 + 15n - 30n^2 + 15n^3 - 2n^4) + 4(22 - 5n\right.\\ 
 &\phantom{=} + 16n^2 - 20n^3 + 9n^4 - 2n^5)\epsilon - 4(12 + 16n - 21n^2 + 2n^3 + 5n^4 - 2n^5)\epsilon^2\\
 &\phantom{=} \left.+ n(24 - 38n + 25n^2 - 7n^3)\epsilon^3\right]
\end{splitequation}
and
\begin{equation}
 \mathi \expect{H^{(1)}(x)S^{(1)}_{\textrm{G}, U}}_0 = -H^{n - 1}C_{\mathrm{G}, U}(n,\epsilon)\eqend{,}
\end{equation}
with
\begin{splitequation}\label{eq:C_G_U}
 C_{\mathrm{G}, U}(n,\epsilon) 
 & = -\frac{(1 - \epsilon)^{n - 4}(2 - \epsilon)A^{(n)}_\mu}{128(n - 2)(n - 1)(n + 2)\epsilon}\left[32(n - 1)(11 - 13n - 5n^2 + 4n^3)\right.\\ 
 & \phantom{=} - 8(-30 + 311n - 222n^2 - 65n^3 + 50n^4 - 3n^5 + n^6)\epsilon + 4(48 + 598n \\
 & \phantom{=} + 555n^2 - 94n^3 + 105n^4 - 13n^5 + 5n^6)\epsilon^2 - 2(64 + 336n - 370n^2 - 71n^3\\ 
 & \phantom{=} \left. + 56n^4 - 5n^5 + 6n^6)\epsilon^3 + n(n + 2)(32 - 32n - 11n^2 + 6n^3 + n^4)\epsilon^4\right]\eqend{.}
\end{splitequation}
The renormalised expectation value for $\mathcal{H}$ then yields
\begin{equation}\label{eq:invariant_H_renormalised_final_cte_epsilon}
 \langle\mathcal{H}_\mathrm{ren}(x)\rangle = H + \kappa^2\epsilon H^3\ln a\lim_{n \to 4}[(n - 4)C(n,\epsilon)]\eqend{,}
\end{equation}
where
\begin{equations}[eq:C_total]
 C(n, \epsilon) & \equiv C_1(n,\epsilon) + C_2(n,\epsilon)\eqend{,}\\
 C_1(n,\epsilon) & \equiv C_\mathrm{GH}(n,\epsilon) + C_{\mathrm{G},V}(n,\epsilon) + C_{\mathrm{G},U}(n,\epsilon)\eqend{.}
\end{equations}

In the case of a matter-dominated universe, we have $\epsilon_\mathrm{matt} = \frac{n - 1}{2}$, 
\begin{equation}
 C(n,\epsilon_{\mathrm{matt}}) = -\frac{1}{n - 4}\frac{229}{192\pi^2} + \bigo{(n - 4)^0}
\end{equation}
and
\begin{equation}
 \langle\mathcal{H}_\mathrm{ren}(x)\rangle = H\left(1 - \frac{229}{128\pi^2}\kappa^2H^2\ln a\right)\eqend{.}
\end{equation}
In a radiation-dominated universe, we have $\epsilon_\mathrm{rad} = \frac{n}{2}$,
\begin{equation}
 C(n,\epsilon_{\mathrm{rad}}) = 0
\end{equation}
and
\begin{equation}
 \langle\mathcal{H}_\mathrm{ren}(x)\rangle = H\eqend{.}
\end{equation}
Finally, in the case of small $\epsilon$ we find
\begin{equation}
 \langle\mathcal{H}_\mathrm{ren}(x)\rangle = H\left(1 + \frac{63}{768\pi^2}\kappa^2\epsilon H^2\ln a\right)\eqend{.}
\end{equation}

% the iopart definition
\providecommand\newblock{\ }
\bibliography{references}

\end{document}